\documentclass[lettersize,journal]{IEEEtran}
\usepackage{amsmath,amsfonts}
\usepackage{algorithmic}
\usepackage{algorithm}
\usepackage{array}
\usepackage{textcomp}
\usepackage{stfloats}
\usepackage{url}
\usepackage{verbatim}
\usepackage{graphicx}
\usepackage{cite}
\usepackage{bm}
\usepackage{siunitx}
\usepackage[hang,small,bf]{caption}
\usepackage[subrefformat=parens]{subcaption}
\newcommand{\figref}[1]{Fig.\,\ref{#1}}
\newcommand{\Figref}[1]{Figure\,\ref{#1}}

\newcommand{\secref}[1]{Sec.\,\ref{#1}}

\newcommand{\eref}[1]{Eq.\,(\ref{#1})}

\newtheorem{prob}{Problem}
\newtheorem{remark}{Remark}

\begin{document}

\title{Model Predictive Control of Smart Districts \\ Participating in Frequency Regulation Market: \\
A Case Study of Using Heating Network Storage }

\author{Hikaru Hoshino,~\IEEEmembership{Member,~IEEE,} ~T. John Koo,~\IEEEmembership{Senior Member,~IEEE,}~ Yun-Chung Chu,~\IEEEmembership{Senior Member,~IEEE,}~Yoshihiko Susuki,~\IEEEmembership{Member,~IEEE}
\thanks{Manuscript received April XX, 20XX; revised August XX, 20XX. This work was supported in part by Grant-in-Aid for Scientific Research (KAKENHI) from the Japan Society for
Promotion of Science (\#23K13354).}%
\thanks{H. Hoshino is with the Department of Electrical Materials and Engineering, University of Hyogo, 2167 Shosha, Himeji, Hyogo, 671-2280, Japan (email: hoshino@eng.u-hyogo.ac.jp). }%
\thanks{T.~J.~Koo and Y.-C.~Chu are with Hong Kong Applied Science and Technology Research Institute (ASTRI), 2 Science Park East Avenue, Hong Kong Science Park, Hong Kong (email: johnkoo@astri.org; ycchu@astri.org). }%
\thanks{Y.~Susuki is with the Department of Electrical Engineering, Kyoto University, Katsura, Nishikyo, Kyoto, 615-8510, Japan (email: susuki.yoshihiko.5c@kyoto-u.ac.jp). }%
}

\markboth{}%
{Shell \MakeLowercase{\textit{et al.}}: A Sample Article Using IEEEtran.cls for IEEE Journals}

\IEEEpubid{~}

\maketitle

\begin{abstract}
Flexibility provided by Combined Heat and Power (CHP) units in district heating networks is an important means to cope with increasing penetration of intermittent renewable energy resources, and various methods have been proposed to exploit thermal storage tanks installed in these networks. This paper studies a novel problem motivated by an example of district heating and cooling networks in Japan, where high-temperature steam is used as the heating medium. In steam-based networks, storage tanks are usually absent, and there is a strong need to utilize thermal inertia of the pipeline network as storage. However, this type of use of a heating network directly affects the operating condition of the network, and assuring safety and supply quality at the use side is an open problem. To address this, we formulate a novel control problem to utilize CHP units in frequency regulation market while satisfying physical constraints on a steam network described by a nonlinear model capturing dynamics of heat flows and heat accumulation in the network. Furthermore, a Model Predictive Control (MPC) framework is proposed to solve this problem. By consistently combining several nonlinear control techniques, a computationally efficient MPC controller is obtained and shown to work in real-time. 
\end{abstract}


\section{Introduction}

\IEEEPARstart{W}{ith} increasing penetration of renewable energy sources and resultant closing down of existing thermal power generation units,  the need for flexible and controllable generation is increasing to maintain operational security and reliability of electric power grids \cite{bollen11}. 
One of the solutions for this is to exploit the inherent operational flexibility of Combined Heat and Power (CHP) units \cite{ostergaard10,korpela17,mugnini21}. 
Recently,  within deregulated electricity market environments,  many types of demand-side resources have begun to participate in ancillary services markets \cite{du20,abbas21}, and CHP units are also expected to participate in these markets \cite{mahani20,gao22}.  

%
Most CHP units are used in district heating networks.
Since the total heat production needs to be balanced with the total heat consumption, this can impose a limitation to the long-term power production. 
However,  the heat load and the power production can be temporarily decoupled by using thermal storage in heating networks. 
Since 
sensible thermal storage tanks are 
commonly installed next to CHP units in hot-water based heating networks in some European countries such as Sweden and Denmark \cite{hennessy19},  
many operation and control methods have been proposed for integrated electricity and heating networks equipped with storage tanks \cite{parisio17,cao19,zhang19,li22,good19, martinez19, long19}. 

\IEEEpubidadjcol

\begin{figure}[!tb]
\centering
\includegraphics[width=0.95\linewidth]{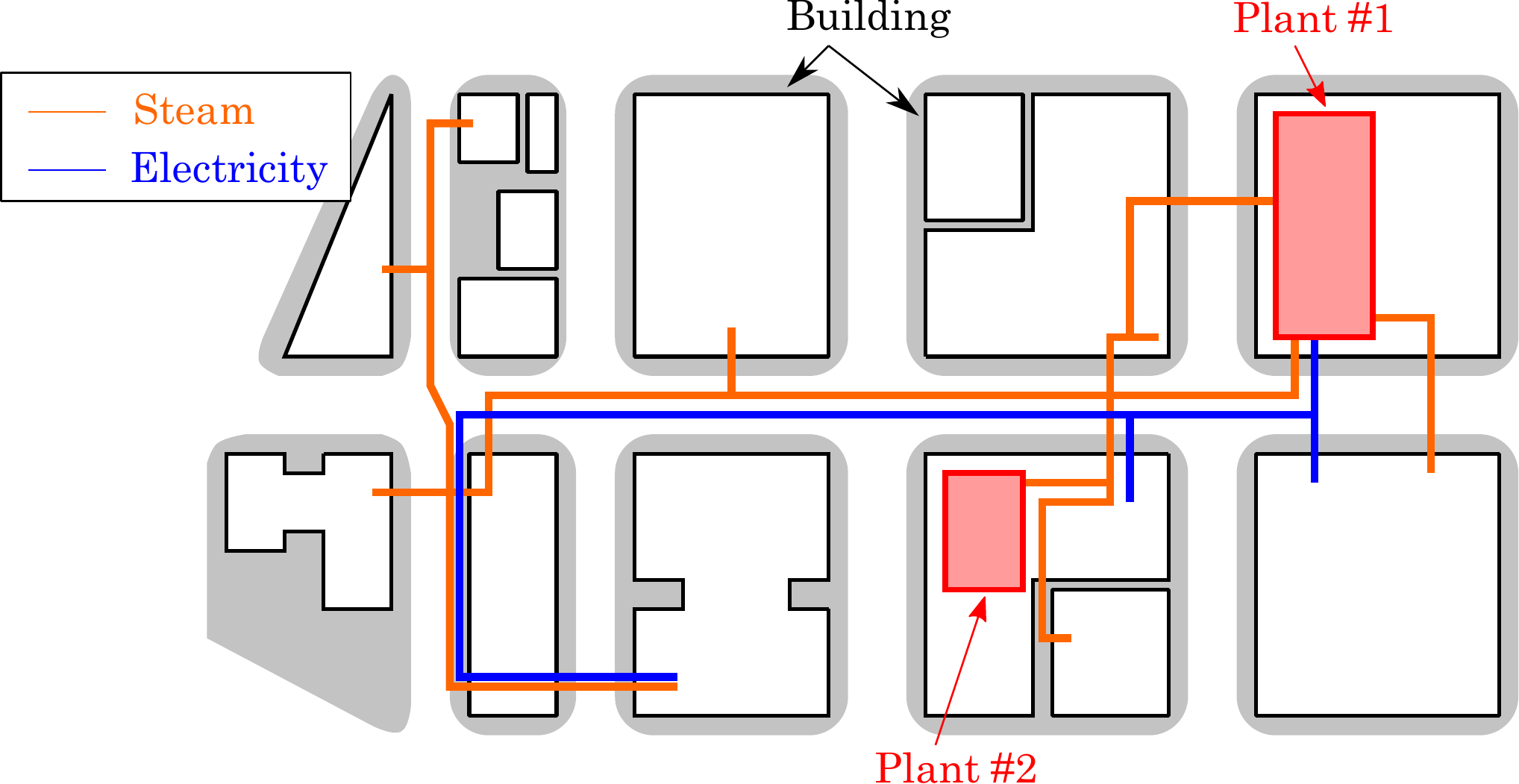}
\caption{A practical example of DHC system in an urban district in Tokyo. }
\label{fig:practical_architecture}
\end{figure}

In this paper,  we study a novel control problem motivated by a practical example of steam networks for District Heating and Cooling (DHC) of urban districts in Japan  (see \figref{fig:practical_architecture}). 
Although low-temperature hot-water networks are preferred in European countries to decrease distribution losses \cite{lund14,werner17}, high-temperature steam is widely used in urban commercial/business districts in some Asian countries \cite{suzuki21,zeng21}, because the distribution loss is relatively low due to dense demand and small supply area, and high-efficiency steam-driven chillers are available for cooling.   
In these networks, thermal storage tanks are usually absent, and thus there is a strong need for utilizing \emph{network storage} \cite{hennessy19} that comes from thermal inertia of the contents of the pipeline network and does not require a large investment cost for e.g. a steam accumulator. 
However, this type of use of a heating network directly affects the operating condition of the network, and 
assuring safe operation and supply quality at the user side is an open problem to be studied. 
To address this, we formulate and solve a novel control problem to enable participation of CHP units in frequency regulation market while satisfying physical constraints on a steam network using a nonlinear dynamical model describing heat flows and heat accumulation in the network.

\subsection{Related Literature}

Related to this work, energy systems integration \cite{omalley13,wu16} and multi-energy systems \cite{mancarella14,chertkov20} have attracted a lot of interest in power and energy research community. 
As a starting point, the \emph{energy hub} concept was presented and formulated by Geidl and Andersson \cite{geidl07:IEEETPS, geidl07:PESMAG}, where they provided a modeling framework intended for the optimal power flow problem among multiple distinct energy infrastructures, including natural gas, electricity, and district heating networks.
With the energy hub concept, dispatch factors and matrix models are introduced to represent energy conversions between different energy carriers, and thereby various methods have been presented for operation, planning, and, design optimizations in steady-state conditions. 
Time-dependent variables or parameters were not taken into consideration.

Based on the formulation of the energy hub concept,  a Model Predictive Control (MPC) approach has been presented in \cite{arnold09}, where energy storage systems are included in the model, and the time dependency of the state of charges of these storage systems are taken the into consideration. 
After that, various operational optimization methods have been reported in e.g. \cite{good19, martinez19, long19, parisio17,cao19,zhang19,li22}, postulating that energy flows can be modeled by using static formulations, while dynamic characteristics are considered for storage systems.  
In particular, provision of flexibility from multi-energy systems have been discussed in \cite{good19, martinez19,long19}. 
However,  these works do not consider the contents of the pipeline networks as storage, and for this purpose, more detailed modeling is required to represent the dynamic characteristics of district heating networks. 
Although some papers have considered dynamic characteristics of district heating networks \cite{pan16,pan19}, their  objective is on security assessment of the networks, and provision of flexibility has not been explored.

%

%
Furthermore, while most of the existing works mainly focus on the operational optimization on the time scale of tens of minutes to hours, from the viewpoint of frequency regulation, it is also necessary to study on a control problem on a shorter time scale. 
Regarding this,  experimental validation tests have been reported in \cite{korpela17} using municipal coal fired CHP plants located in Finland. 
To test technical feasibility of using CHP units for a short-term frequency regulation,
dynamic responses of CHP units and district heating networks are analyzed by measured data obtained through step-response tests based on requirements for attending Automatic Frequency Restoration Reserve (FRR-A) in Nord Pool. 
As a result, the desired FRR-A requirements were fulfilled implying CHP units can  contribute to the balancing of the power grid. 
However, it was reported that significant fluctuations in live steam pressure were observed due to  slow response of coal mills and  interference of control loops. 
Thus, it was concluded in \cite{korpela17} that in addition to analyzing the capability of CHP units to adjust electricity production, the dynamic characteristics of other components and controllers should also be considered. 
However, to the best of our knowledge, no consistent control approach has been reported to enable participation of CHP units in a short-term frequency regulation while taking the dynamic characteristics of heating networks into consideration.

\subsection{Statement of Contributions}

The first major contribution of this paper is to formulate a novel control problem to enable participation of CHP units in frequency regulation market by exploiting thermal network storage.  
To ensure safe operation and supply quality at the user side, 
we consider a nonlinear dynamical model of heat flows and heat accumulation in steam networks developed in our previous studies \cite{hoshino14:nolta,hoshino16,hoshino19}, whereas most of the existing works on MPC design for multi-energy systems postulate static energy flows. 
Furthermore, the primary objectives of the aforementioned papers are related to operational optimization on the time scale of tens of minutes to hours.  
In this paper, we aim to design an MPC controller that works on the time scale of seconds to minutes 
to enable the participation of CHP units in Automatic Generation Control (AGC) \cite{kundur94}, where the contributing resources are required to  adjust their outputs in response to a dispatch signal generated by the grid operator. 
Specifically, PJM Regulation D (Reg-D) signal \cite{pjm_manual12} is considered as a typical AGC signal, and an output tracking control problem is formulated to maximize tracking performance for Reg-D signal, while satisfying specified physical constraints to ensure safety and supply quality of the steam network.

The second major contribution is to propose a computationally efficient MPC framework considering nonlinear dynamics of integrated electricity and steam networks. 
By using a nonlinear model to predict the future evolution of the system, the resulting MPC controller usually requires a nonlinear optimization problem to be solved at each time step, which can be the computational burden in the implementation. 
In this paper, based on a structural analysis of the dynamical model mentioned above,  
we establish an appropriate MPC framework by consistently combining three techniques of 1) coordinate transformation based on the geometric nonlinear control theory \cite{isidori95,sastry99} to decompose the entire dynamics into \emph{linear outer} dynamics related to the tracking performance and \emph{nonlinear internal} dynamics related to safety and supply quality of the steam network,  2)  successive linearization scheme based on Iterative Linear Quadratic Regulator (ILQR) \cite{li04,todorov05} to derive a Linear Time-Varying (LTV) system approximating the nonlinear internal dynamics, and 3)  a LTV-MPC controller to calculate the control input.
It is shown that the obtained MPC controller achieves more than ten times faster computational speed than a standard nonlinear MPC controller and can be used in real-time with the sampling period of two seconds, which is the same as the resolution of the Reg-D signal.

The third major contribution is to provide detailed discussions on the behavior of the proposed MPC controller. 
For this purpose, we introduce a benchmark system that captures essential features of the DHC network in an urban commercial/business district in the central Tokyo, Japan. 
In this benchmark system, two CHP units are operated in different sites (DHC plants) connected via electricity and steam networks. 
It will be shown that the complicated behaviors of the proposed MPC controller can be well understood by using a two-dimensional portrait of energy flows in the two-site benchmark system.

%

\section{Problem Statement} \label{sec:model}

In this section, we introduce the physical architecture of the benchmark system studied in this paper in \secref{sec:system_description}, present its nonlinear state-space model  in \secref{sec:state_space_model}, and formulate the control problem addressed in this paper in \secref{sec:control_objectives}.  
%
%

\subsection{Physical Architecture} \label{sec:system_description}

The benchmark system studied in this paper is devised based on a practical example of District Heating and Cooling (DHC) networks seen in urban districts in Japan. 
\Figref{fig:practical_architecture} shows a schematic diagram of electricity and steam networks in Yurakucho district \cite{suzuki21}, which is located  between Tokyo Station and the Imperial Palace in the central Tokyo. 
In this example, high-temperature steam is supplied to eleven buildings from two DHC plants located in a spatially distributed manner. 
Although low-temperature hot-water networks are preferred for district heating in European countries to decrease distribution losses \cite{lund14,werner17}, high-temperature steam is widely used in urban commercial/business districts in Japan. 
This is because high-efficiency steam-driven chillers are available for cooling, and the distribution loss is lower due to dense demand and smaller supply area than European heating networks.  
In each DHC plant, a CHP unit is installed to supply electricity as well as steam. 
Although the main reason for installing  CHP units has been to improve energy efficiency, in recent years, more attention has been focused on the security of electricity supply  in the event of a disaster like an earthquake \cite{suzuki21}. 
With this reason,  private electricity distribution lines are installed to supply electricity to five buildings in case  of the interruption of electricity supply from the commercial power grid. 
Thus, this system can be understood as a Microgrid \cite{hatziargyriou07} integrated with a steam supply network.

\begin{figure}[!t]
 \centering
 \includegraphics[width=0.9\hsize]{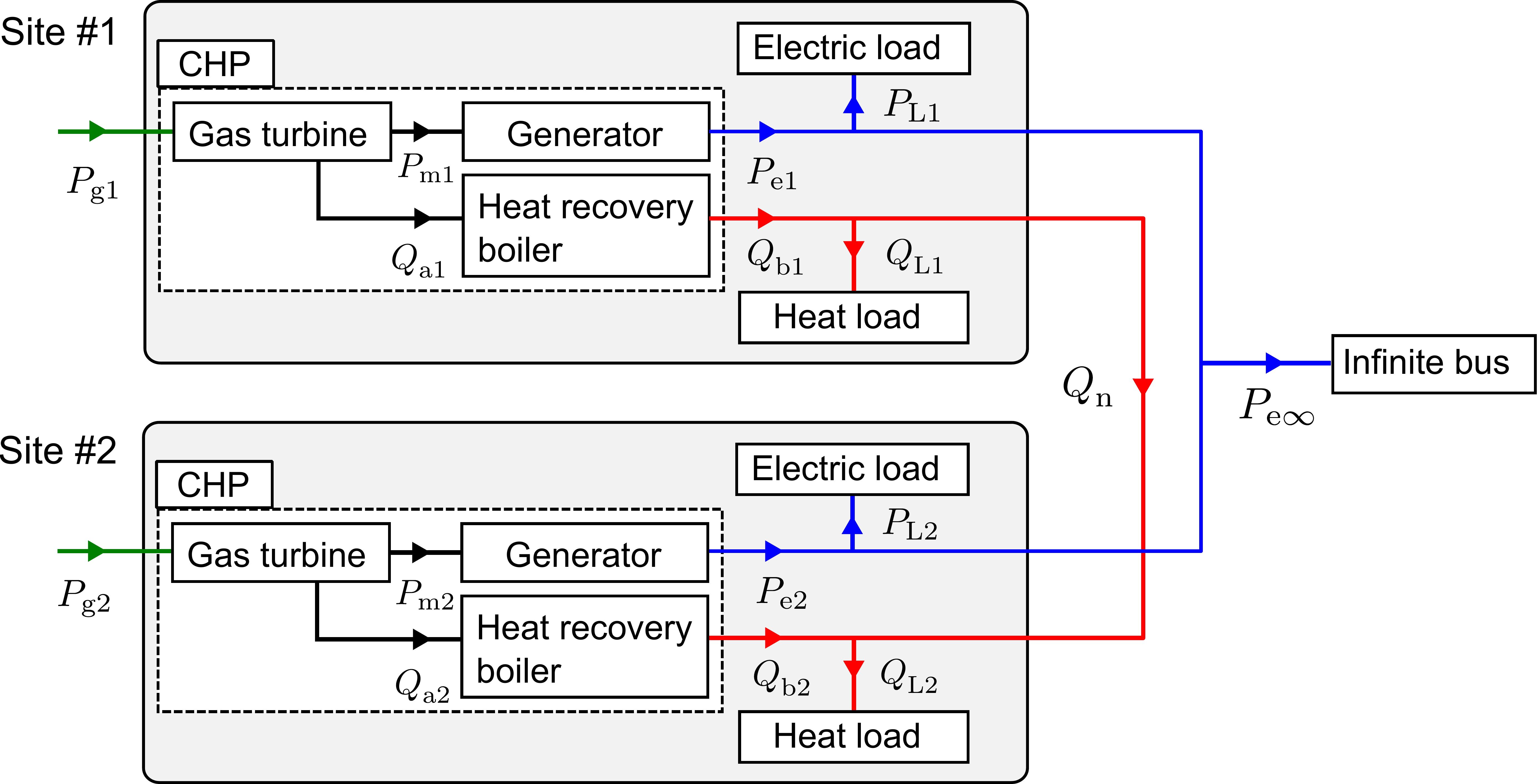}
 \caption{%
 Energy flow diagram of the two-site system. 
 The arrows show the positive directions of the energy flows.}
 \label{fig:twosite_system}
\end{figure}

Based on this example, this paper studies a two-site benchmark system as shown in \figref{fig:twosite_system}. 
The concept of \emph{site} stands for a unit of energy system that consists of a CHP unit, an electric load, and a heat load. 
Each CHP unit consists of a gas turbine, a synchronous generator, and a heat recovery boiler.
The two sites are electrically interconnected through a distribution network and are also connected to a commercial power grid, which can be modeled as an infinite bus \cite{kundur94}. 
In the context of power system engineering, the two-site system (with the external infinite bus) can be regarded as an extension of the so-called three-node network, for which in-depth analysis has been done on static and dynamic characteristics in e.g.  \cite{araposthatis81,arapostathis83,hasegawa99}. 
In terms of the heat supply, the two-site configuration is minimal for considering bidirectional heat transfer  \cite{nedo_kobe18} between different sites distributed in an urban district. 
The arrows in the figure show energy flows in the system. 
The fuel flow $P_{{\rm g}i}$ to the gas turbine at site $\# i$, for $i=1,\,2$,  can be a control input of this system, and as a result of fuel combustion, the mechanical power $P_{{\rm m}i}$ is produced and then converted to the electric power $P_{{\rm e}i}$ by the synchronous generator in each CHP unit. 
The power $P_{{\rm e}i}$ is partly consumed by the electric load $P_{\mathrm{L}i}$ and supplied to the infinite bus $P_{{\rm e}\infty}$ (to the commercial power grid). 
Also, from the exhaust gas of the gas turbine $\# i$, the heat flow $Q_{{\rm a}i}$ is absorbed by the heat recovery boiler, and the generated steam is supplied from the CHP unit with the heat flow rate $Q_{{\rm b}i}$. 
It is partly consumed by the heat load $Q_{\mathrm{L}i}$ at each site $\# i$, and bidirectional heat transfer occurs between the two sites with the heat transfer rate $Q_\mathrm{n}$. 
Furthermore, although not explicitly shown in the figure, excess or deficit heat is accumulated to or decumulated  from the entire steam network.

\subsection{State-space Model} \label{sec:state_space_model}

A state-space model of the two-site benchmark system has been developed in our previous work \cite{hoshino14:nolta,hoshino16,hoshino19} for describing the dynamics evolved  on the time-scale of seconds to minutes. 
The entire system consists of three subsystems and can be described in the following form:%
\begin{subequations} \label{eq:system_equation}
\begin{align}
 \underbrace{
 \begin{bmatrix} \dot{\bm{x}}_{\rm g} \\ \dot{\bm{x}}_{\rm e} \\ \dot{\bm{x}}_{\rm h} \end{bmatrix}
 }_{\dot{\bm{x}}}
 & =
 \underbrace{
    \begin{bmatrix} 
       \bm{f}_{\rm g}(\bm{x}_{\rm g}) \\ \bm{f}_{\rm e}(\bm{x}_{\rm e},\,\bm{x}_{\rm g}) \\ \bm{f}_{\rm h}(\bm{x}_{\rm h},\,\bm{x}_{\rm g}) 
    \end{bmatrix}
}_{\bm{f}(\bm{x})}
+  
\sum_{i=1}^2
\underbrace{
   \begin{bmatrix} \bm{g}_{{\rm g}i}(\bm{x}_{\rm g}) \\ 0  \\ 0  \end{bmatrix} 
}_{\bm{g}_i(\bm{x})}
  u_i, 
 \\
 \bm{y} & = 
\underbrace{
   \begin{bmatrix} P_{\mathrm{e}\infty}(\bm{x}_\mathrm{e}) & Q_\mathrm{n}(\bm{x}_\mathrm{h}) & R(\bm{x}_\mathrm{h}) \end{bmatrix}^\top
}_{\bm{h}^\top(\bm{x})} \label{eq:outputs}
\end{align}
\end{subequations}
where $\bm{x}_{\rm g} \in \mathbb{R}^6$ stands for the state variable for the gas turbines, $\bm{x}_{\rm e} \in \mathbb{R}^4$ for the state variable for the electric subsystem consisting of the generators interconnected by the distribution network, and $\bm{x}_{\rm h} \in \mathbb{R}^3$ for the state variable for the heat subsystem consisting of the boilers and the steam pipeline network. 
The control inputs $u_1$ and $u_2$ stand for the normalized reference signals for the fuel inputs to the gas turbines  at site $\# 1$ and $\# 2$, respectively. 
While detailed explanations of the these variables and the representations of the functions  $\bm{f}_{\rm g}$, $\bm{f}_{\rm e}$, $\bm{f}_{\rm h}$, $\bm{g}_{\rm g1}$, and $\bm{g}_{\rm g2}$ are given in \cite{hoshino14:nolta,hoshino16,hoshino19},  here we briefly explain the physical meanings of the dynamics described by the models for the three subsystem. 
First, the model of the subsystem for the gas turbines represents the dynamics of fuel valve positioning, fuel distribution and compressor discharging in each gas turbine to describe the time courses of the mechanical power $P_{\mathrm{m}i}$ and the absorbed heat flow $Q_{\mathrm{a}i}$. 
Second, the model of the electric subsystem represents the dynamics of the rotor angles of the generators to describe the time courses of the active power flows $P_{\mathrm{e}i}$ and $P_{\mathrm{e}\infty}$ in the distribution network. 
Finally, the model of the heat subsystem represents the dynamics of steam pressure supplied by the boilers and the steam velocities to describe the nonlinear characteristics of the heat flows ($Q_{\mathrm{b}i}$ and $Q_\mathrm{n}$) and heat accumulation, which will be quantified by the mean pressure $R(\bm{x}_\mathrm{h})$.



\subsection{Model Structure and Control Problem} \label{sec:control_objectives}

Here we discuss the structure of the model \eqref{eq:system_equation} and clarify the control objectives considered in this paper. 
A primary objective is to control the active power $y_1 = P_\mathrm{e\infty}$ flowing into the infinite bus (the commercial power grid) to achieve the tracking of a dispatch signal generated by the grid operator. 
%
%
%
This can be achieved by appropriately adjusting the control inputs $u_1$ and $u_2$ governing the fuel inflows to the gas turbines at site $\# 1$ and $\# 2$, respectively. 
However, this manipulation also affects the steam network as well.  
While a steam network exhibits multiple time-scale responses \cite{hoshino16}, in the short-term regime,  the steam velocity (the heat flow) inside the pipeline can deviate from the planned value due to the above manipulation. 
If the steam velocity unintentionally drops to a value near zero, it may cause a condensation-induced water hammer due to steam stagnation  \cite{zhong15}, which will threaten the safety of the whole heating network.  
Thus, it is desirable to regulate the heat flow $Q_\mathrm{n}$ (the output $y_2$) by decoupling it from the electric power (the output $y_1$).

Based on the above,  an output tracking control problem can be considered for the output pair $(y_1,\, y_2)$ and the input pair $(u_1, \, u_2)$. 
Then, for an input-affine nonlinear system like the model \eqref{eq:system_equation}, a coordinate transformation can be applied based on  the geometric nonlinear control theory \cite{isidori95,sastry99} to analyze the structure of the controlled system. 
We performed this analysis in \cite{hoshino19} and obtained a coordinate transformation $\bm{\Phi}$ that is a diffeomorphism on a subset $D$ of the state space: 
 \begin{align}  \label{eq:coordinate_transformation}
   \bm{\Phi}:  \bm{x} \mapsto (\bm{\xi}_\mathrm{e}, \,\bm{\xi}_\mathrm{h}, \, \bm{\eta})
 \end{align}
where $\bm{\xi}_\mathrm{e} \in \mathbb{R}^5$ (or $\bm{\xi}_\mathrm{h} \in \mathbb{R}^4$) stands for the output $y_1$ (or $y_2$) and its time derivatives up to the fourth order (or third order) as follows: 
\begin{subequations} \label{eq:state_transformation}
\begin{align}  \bm{\xi}_{\rm e} & := [ h_{\rm 1}(\bm{x}), \, L_f h_{\rm 1}(\bm{x}) ,\, \dots, \, L_f^4h_{\rm 1}(\bm{x}) ]^\top,  \label{eq:xi_e} \\
 \bm{\xi}_{\rm h} & := [h_{\rm 2}(\bm{x}), \, L_f h_{\rm 2}(\bm{x}),  \,  L_f^2h_{\rm 2}(\bm{x}),  \,  L_f^3h_{\rm h}(\bm{x}) ]^\top, 
\end{align}
\end{subequations}
where $L_f {h}$ stands for the Lie derivative of a function $h$ along the vector field $\bm{f}$. 
 With this transformation, the model \eqref{eq:system_equation} can be translated into the following form \cite{hoshino19}: 
\begin{subequations}
\label{eq:linearized_system}
 \begin{align}
 &\dot{\bm{\xi}}_\mathrm{e} = \mathsf{A}_\mathrm{e} \bm{\xi}_\mathrm{e} + \mathsf{B}_\mathrm{e} v_1, \label{eq:xi_e_subsystem} \\ 
 &\dot{\bm{\xi}}_\mathrm{h} = \mathsf{A}_\mathrm{h} \bm{\xi}_\mathrm{h} + \mathsf{B}_\mathrm{h} v_2, \label{eq:xi_h_subsystem} \\
 &\dot{\bm{\eta}} = \bm{q}(\bm{\xi}_\mathrm{e}, \, \bm{\xi}_\mathrm{h}, \, \bm{\eta}) \label{eq:internal_dynamics}
\end{align}
\end{subequations}
where the dynamics of $\bm{\xi}_{\rm e}$ and $\bm{\xi}_{\rm h}$ are linear and represented by some matrices $\mathsf{A}_\mathrm{e}$, $\mathsf{A}_\mathrm{h}$, $\mathsf{B}_\mathrm{e}$, and $\mathsf{B}_\mathrm{h}$ with virtual inputs $v_1$ and $v_2$ defined in the procedure of the transformation. 
The variable $\bm{\eta} \in \mathbb{R}^4$ stands for the internal state that does not affect the dynamics of the outputs $y_1=\xi_\mathrm{e1}$ and $y_2=\xi_\mathrm{h1}$.

\begin{figure}[!tb]
\begin{minipage}[b]{\linewidth}
\centering
	\includegraphics[width=0.94\linewidth]{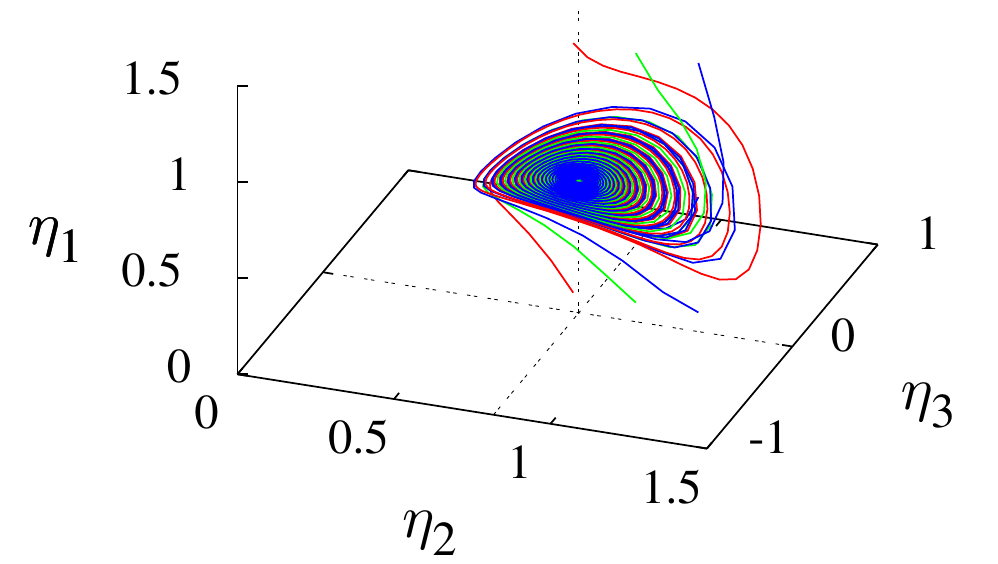}
	\subcaption{Trajectories projected to $(\eta_1,\, \eta_2,\, \eta_3)$ space. } 
	 \label{fig_sub1} 
	 \vspace{3mm}
	\includegraphics[width=0.94\linewidth]{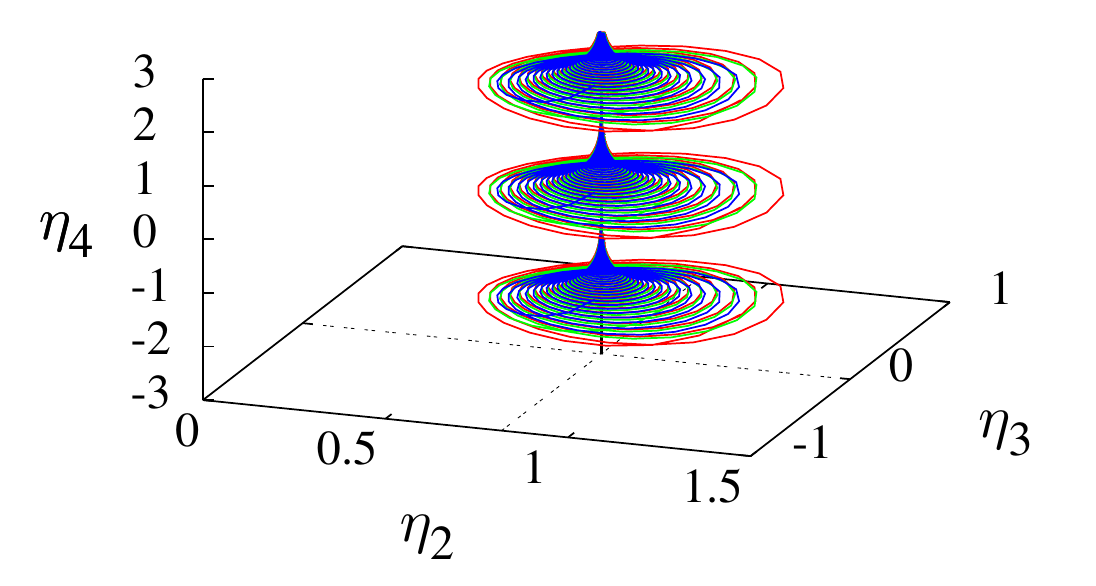}
	\subcaption{Trajectories projected to $(\eta_2,\, \eta_3,\, \eta_4)$ space. }
	 \label{fig_sub1}
\end{minipage}
\caption{Trajectories of the zero dynamics \cite{hoshino19}. }
\label{fig:zero_dynamics}
\end{figure}

Based on the system representation \eqref{eq:linearized_system}, the output tracking or regulation of $y_1$ and $y_2$ can be achieved by using e.g. a linear MPC controller to calculate the virtual inputs $v_1$ and $v_2$ and converting them to the original control inputs $u_1$ and  $u_2$. 
Then, if the internal dynamics have an asymptotically stable equilibrium point, such a system is called a minimum-phase system, and the perfect tracking can be achieved while ensuring the boundedness of the entire state. 
However, it has been confirmed in \cite{hoshino19} that the system \eqref{eq:system_equation} is a non-minimum phase system, and the internal dynamics \eqref{eq:internal_dynamics} have no asymptotically stable equilibrium point. 
\Figref{fig:zero_dynamics} shows the trajectories of the so-called zero dynamics, which are the internal dynamics when the outputs $y_1$ and $y_2$ are kept constant for all the time. 
While the states $\eta_1$ to $\eta_3$, which are associated with the gas turbine and the electric subsystem, exhibit convergent behavior,  the state $\eta_4$ is not converging to a constant value. 
The state $\eta_4$ stands for the steam pressure $y_3 =  R(\bm{x}_\mathrm{h})$,  and its dynamics represents the heat accumulation to  the pipeline network. 
To ensure supply quality at the user side, the pressure $y_3 = \eta_4$ must be kept within an acceptable range. 
Thus, in this paper, we consider the following output tracking control problem, where the perfect tracking will be compromised when the priority should be given to the safety and the supply quality of the steam network. 


%

\begin{prob} \label{prob:tracking}
Consider the system described by the model \eqref{eq:system_equation}. 
Design a controller that maximizes tracking performance of the output $y_1$ to an external reference signal, while achieving regulation of the output $y_2$ and  ensuring the boundedness of the output $y_3$ within stipulated upper and lower limits by using the control inputs $u_1$ and $u_2$ which may also have upper and lower limits.  
\end{prob}

\section{Proposed MPC Framework} \label{sec:controller_design}

In this section, we propose an appropriate MPC framework for the control problem introduced above. 
The overview of the proposed framework 
is explained in \secref{sec:state_transformation}, and the successive linearization scheme to derive an approximate LTV system is described in \secref{sec:linerization}. Finally, the MPC formulation is presented in \secref{sec:mpc_formulation}. 


\subsection{Overview} \label{sec:state_transformation}

The proposed MPC framework is based on the coordinate transformation \eqref{eq:coordinate_transformation}. 
With this transformation, the original model can be decomposed into the \emph{linear outer} dynamics related to the tracking performance of $y_1$ and $y_2$ given by \eqref{eq:xi_e_subsystem} and \eqref{eq:xi_h_subsystem} and the \emph{nonlinear internal} dynamics \eqref{eq:internal_dynamics} describing the time course of the pressure $y_3$ related to the supply quality of the steam network. 
To solve the Problem 1, we can use an objective function that related only to the outer dynamics to maximize the tracking performance for $y_1$ and $y_2$ as much as possible. 
Then, the boundedness of the pressure $y_3$ can be mediated by imposing a constraint on the internal dynamics.

\begin{remark}
Similar control methods based on the combination of a nonlinear coordinate transformation and a linear MPC controller have been proposed  in \cite{nevistic94,nevistic96,margellos10,simon13,schnelle15} and applied to several power and energy applications \cite{gionfra16,quan21,kong23}. However, in these previous studies, the class of controlled systems was limited to exactly linearizable systems, where an entire system can be transformed to a linear system without a nonlinear part, or minimum-phase systems mentioned above. 
In this paper, we propose an MPC framework that can be used for a non-minimum phase system by further combining this technique with   a successive linearization scheme based on the Iterative Linear Quadratic Regulator (ILQR) \cite{li04, todorov05} as presented below. 

\end{remark}

\subsection{Discretization and Linearization of Internal Dynamics} \label{sec:linerization}

Before presenting the MPC formulation, discretization of the model is performed. 
By applying the zero-order hold (ZOH) discretization to the model \eqref{eq:linearized_system} at the sampling period $T_\mathrm{s}$, the following discrete time system can be defined: for $\bm{z}[k]:=[\bm{\xi}_\mathrm{e}(kT_\mathrm{s})^\top, \, \bm{\xi}_\mathrm{h}(kT_\mathrm{s})^\top ]^\top$, $\bm{w}[k] := \bm{\eta}(kT_\mathrm{s})$, and $\bm{v}[k] := \bm{v}(k T_\mathrm{s})$, 
\begin{subequations}
\begin{align}
 &\bm{z}[k+1] = \mathsf{F} \bm{z}[k] + \mathsf{G} \bm{v}[k],  \label{eq:linear_subsystem} \\
 &\bm{w}[k+1] = \bm{\Psi}( \bm{z}[k], \bm{w}[k], \bm{v}[k])  \label{eq:discrete_zero_dynamics} 
 \end{align}
\end{subequations}
where the matrices $\mathsf{F}$ and $\mathsf{G}$ are given by 
\begin{align}
&\mathsf{F} :=  \exp \left(  \begin{bmatrix} \mathsf{A}_\mathrm{e} & 0 \\ 0 &  \mathsf{A}_\mathrm{h} \end{bmatrix} T_\mathrm{s} \right), \\
&\mathsf{G} := \int_0^{T_\mathrm{s}} \exp \left(  \begin{bmatrix} \mathsf{A}_\mathrm{e} & 0 \\ 0 &  \mathsf{A}_\mathrm{h} \end{bmatrix} \tau  \right) \begin{bmatrix} \mathsf{B}_\mathrm{e} & 0 \\ 0 & \mathsf{B}_\mathrm{h} \end{bmatrix} \mathrm{d}\tau, 
\end{align}
and can be easily computed. 
The transition map $\bm{\Psi}$ for the internal dynamics is defined by
\begin{align}
&\bm{\Psi} (\bm{z}[k], \bm{z}_\mathrm{\eta}[k], \bm{v}[k]) \notag  \\
& := \bm{z}_\mathrm{\eta}[k] + \int_{kT_\mathrm{s}}^{ (k+1)T_\mathrm{s} } \bm{q}(\bm{\xi}_\mathrm{e}(\tau), \, \bm{\xi}_\mathrm{h}(\tau), \, \bm{\eta}(\tau)) d\tau \label{eq:def_transition_map}
\end{align}
where the trajectories $\bm{\xi}_\mathrm{e}(\tau)$ $\bm{\xi}_\mathrm{h}(\tau)$, and $\bm{\eta}(\tau)$ are given by integrating the dynamics \eqref{eq:linearized_system} from the initial conditions at $\tau = k T_\mathrm{s}$ with a constant input $\bm{v}(\tau) \equiv \bm{v}[k]$. 

Since the map  $\bm{\Psi}$ is not available in a closed form, this paper applies  a linearization scheme inspired by the ILQR. 
At each time step $k$,  
we have the information on the sequence of the control action $\{ \hat{\bm{v}}[k+l|k-1] \}_{l=0}^{N-2}$ planned at the previous step $k-1$, and the current state $(\bm{z}[k],\, \bm{w}[k])$. 
Thus, we can compute a nominal trajectory, represented by  $\{ (\bar{\bm{z}}[k+l|k], \bar{\bm{w}}[k+l|k])  \}_{l=0}^{N-1}$. 
Then, by considering the linearized dynamics of the equation \eqref{eq:discrete_zero_dynamics} around the nominal trajectory,  
we have the following state equation for $\bm{w}$:
for $l= 1,\,\dots,\,N-1$, 
\begin{align}
 & \hspace{-3mm} \bm{w}[k+l+1]  \notag \\
 = & \bm{\Psi}( \bm{z}[k+l], \bm{w}[k+l], \bm{v}[k+l])  \notag \\
 \approx 
 & \bm{\Psi}( \bar{\bm{z}}[k+l | k],  \bar{\bm{w}}[k+l| k], \hat{\bm{v}}[k+l | k-1]) \notag  \\
    & +\mathsf{F}_z[k+l|k] (\bm{z}[k+l]  - \bar{\bm{z}}[k+l | k]) \notag \\
    & +\mathsf{F}_w[k+l|k] (\bm{w}[k+l]  - \bar{\bm{w}}[k+l | k]) \notag \\
    & +\mathsf{G}_w[k+l|k] (\bm{v}[k+l]  - \hat{\bm{v}}[k+l | k-1])   \label{eq:linearization_procedure}
\end{align}
where $\mathsf{F}_z[k+l|k], \mathsf{F}_w[k+l|k]$, and $\mathsf{G}_w[k+l|k]$ are given by 
\begin{align} \label{eq:derivatives}
 & \hspace{-5mm} ( \mathsf{F}_z[k+l|k], \mathsf{F}_w[k+l|k],  \mathsf{G}_w[k+l|k] )  := \notag \\ 
  \biggl(& \dfrac{ \partial \bm{\Psi} }{ \partial \bm{z} }( \bar{\bm{z}}[k+l|k], \bar{\bm{w}}[k+l|k], \hat{\bm{v}}[k+l|k-1] ), \notag \\
 &  \dfrac{ \partial \bm{\Psi} }{ \partial \bm{w} }( \bar{\bm{z}}[k+l|k], \bar{\bm{w}}[k+l|k], \hat{\bm{v}}[k+l|k-1] ), \notag \\
 &  \dfrac{ \partial \bm{\Psi} }{ \partial \bm{v} }( \bar{\bm{z}}[k+l|k], \bar{\bm{w}}[k+l|k], \hat{\bm{v}}[k+l|k-1] )  \biggr).
\end{align}
Note that the matrices $ \mathsf{F}_z[k+l|k], \mathsf{F}_w[k+l|k]$, and $\mathsf{G}_w[k+l|k]$ 
can be obtained by solving the variational equation \cite{parker89}, which describes the relationship between the Jacobian of $\bm{q}$ in \eqref{eq:internal_dynamics} and the linearization of the map $\bm{\Psi}$. 
Thus, the linearized system can be written as 
\begin{align}
 \bm{w}[k+l+1] = & \mathsf{F}_w[k+l|k] \bm{w}[k+l]  \notag \\ 
 & +  \mathsf{F}_\mathrm{z}[k+l|k] \bm{z}[k+l]  \notag \\ 
 & + \mathsf{G}_w[k+l|k] \bm{v}[k+l]   \notag \\
 & +  \tilde{\bm{w}}[k+l+1|k]
\end{align} 
where $\tilde{\bm{w}}[k+l+1|k]$ is given by 
\begin{align}
\tilde{\bm{w}}[k+l+1|k] : =  & \bar{\bm{w}}[k+l+1|k] \notag \\
   & -\mathsf{F}_z[k+l|k] \bar{\bm{z}}[k+l | k] \notag \\
   & -\mathsf{F}_w[k+l|k]  \bar{\bm{w}}[k+l | k]  \notag \\
   & -\mathsf{G}_w[k+l|k]  \hat{\bm{v}}[k+l | k-1]. 
\end{align}
Note that after the linearization, the system is represented as an LTV system and therefore the optimization problem to compute the control action  $\{ \hat{\bm{v}}[k+l|k] \}_{l=0}^{N-1}$ is much easier than for a nonlinear system.

%
%

\subsection{MPC Formulation} \label{sec:mpc_formulation}

Based on the representation obtained above, here we present the MPC formulation proposed in this paper. 
For a given desired output trajectory $\bm{s}_\mathrm{d}[k]$, let $\bm{z}_\mathrm{d}[k]$ denote the associated desired state trajectory of the linear subsystem \eqref{eq:linear_subsystem},  and  $\bm{v}_\mathrm{d}[k]$ the desired input trajectory. 
Then, the cost function for the tracking MPC can be written as 
\begin{align} \label{eq:cost_function}
 J[k] =& \| \hat{\bm{z}}[k+N|k] - \bm{z}_\mathrm{d}[k+N]  \|^2_{ \mathsf{P}_{z} } \notag \\
  &+ \sum_{l=0}^{N-1} \| \hat{\bm{z}}[k+l|k] - \bm{z}_\mathrm{d}[k+l]  \|^2_{ \mathsf{Q}_{z}} \notag \\
  &+ \sum_{l=0}^{N-1} \| \hat{\bm{v}}[k+l|k] - \bm{v}_\mathrm{d}[k+l]  \|^2_{ \mathsf{R}_{z}}, 
\end{align}
where $\{ \hat{\bm{z}}[k+l|k]  \}_{l=0}^{N}$ and  $\{ \hat{\bm{v}}[k+l|k] \}_{l=0}^{N-1}$ represent the sequences of the state and control action planned at the time step $k$. 
The symbol $\|  \bm{z} \|^2_\mathsf{Q}$ stands for the quadratic form $\bm{z}^\top \mathsf{Q} \bm{z}$ for a vector $\bm{z}$ and a matrix $\mathsf{Q}$. 
The matrix $\mathsf{P}_z$ in the terminal cost is the solution of the algebraic Riccati equation, and $\mathsf{Q}_z$ and $\mathsf{R}_z$ are the weighting matrices for the state and control action, respectively. 
With the cost function \eqref{eq:cost_function}, the proposed MPC algorithm can be stated as follows:

\begin{enumerate}
\item With the current state  $(\bm{z}[k],\, \bm{w}[k])$ and the control sequence $\{ \hat{\bm{v}}[k+l|k-1] \}_{l=0}^{N-2}$ planned at the previous step, calculate the nominal trajectory $\{ (\bar{\bm{z}}[k+l|k], \bar{\bm{w}}[k+l|k])  \}_{l=0}^{N-1}$ by integrating the system \eqref{eq:linearized_system}. 

\item Compute the matrices $ \mathsf{F}_z[k+l|k], \mathsf{F}_w[k+l|k]$, and $\mathsf{G}_w[k+l|k]$  by using the variational equation.

\item Solve the optimization problem given by 
 \begin{subequations}
	\label{eq:optimization_problem}
	\begin{align}
	  & \min_{ \substack{  \{ \hat{\bm{v}}[k+l|k] \}_{l=0}^{N-1} \}_{l=0}^{N}  } } J[k] \\
	&  \hspace{-2mm}  \mathrm{s.t.}   \quad \mathrm{for} \; l=0, 1, \dots, N-1,\notag \\
	 & ( \hat{\bm{z}}[k|k], \hat{\bm{z}}_\mathrm{\eta}[k|k]) = ( \bm{z}[k], \bm{w}[k] ), \\
	  &\hat{\bm{z}}[k+l+1|k] = \mathsf{F} \hat{\bm{z}}[k+l|k] + \mathsf{G} \hat{\bm{v}}[k+l|k], \\
	  &\hat{\bm{w}} [k+l+1|k] =  \mathsf{F}_z[k+l] \hat{\bm{z}}[k+l|k]  \notag \\
	  &\hspace{25mm} + \mathsf{F}_w[k+l] \hat{\bm{z}}_\mathrm{\eta}[k+l|k] \notag \\ 
	  &\hspace{25mm} + \mathsf{G}_w[k+l] \hat{v}[k+l|k] \notag \\ 
	  &\hspace{25mm} +  \tilde{\bm{w}}[k+l+1],  \label{eq:constraint_internal_dynamics} \\
	  & \mathsf{M}_z \hat{\bm{z}}[k+l|k] \le \bm{N}_z, \label{eq:limit_z} \\  
        & \mathsf{M}_w \hat{\bm{w}}[k+l|k] \le \bm{N}_w \label{eq:constraint_w}
	\end{align}
\end{subequations}
\item Execute $\bm{v}[k] = \hat{\bm{v}}[k|k]$ and go to the step 1. 
\end{enumerate}
The matrix $\mathsf{M}_z \in \mathbb{R}^{n_z \times 9}$ (or  $\mathsf{M}_w  \in \mathbb{R}^{n_w \times 4}$) and the vector $\bm{N}_z \in  \mathbb{R}^9$ (or $\bm{N}_w \in  \mathbb{R}^4$) are used for describing inequality constraints on the state $\bm{z}[k]$  (or on the state $\bm{w}[k]$), where $n_z$ (or $n_w$) stands for the number of constraints on  $\bm{z}[k]$ (or on $\bm{w}[k]$). 
For the control problem considered in this paper,  the upper and lower limits of the output $y_3$ can be described by the constraint \eqref{eq:constraint_w}. 
Also, the upper and lower limits of the control inputs $u_1$ and $u_2$ need to be addressed. 
In this paper, we will 
convert the constraints on the inputs to some constraints on the outputs $y_1$ and $y_2$ by considering steady energy flows of the model \eqref{eq:system_equation} as explained in \secref{sec:input_constraints}. 

\section{Simulation results and discussion} \label{sec:simulation}

This section provides simulation results obtained with the MPC controller derived in \secref{sec:controller_design}. 
As an illustrative example, we consider the benchmark system described in \secref{sec:model}.

\subsection{Simulation Settings}

The parameters used for the simulation were taken from \cite{hoshino19}. 
With this parameter setting, the system has an equilibrium point when the reference signals  for the electric power $y_1$ and the heat transfer rate $y_2$ are determined as 
\begin{align}
 ( Y_1^\mathrm{nom}, Y_2^\mathrm{nom}) = (\SI{5.0}{MW}, \SI{1.69}{MJ/s}). 
\end{align}
More detailed description, simulations, and comments can be found in our previous papers \cite{hoshino14:nolta,hoshino16,hoshino19}. 

To examine the ability to participate in power sytem frequency regulation, this paper uses a test data of Regulation D (Reg-D) signal  used in PJM regulation market \cite{pjm_manual12}.   
The test signal is normalized and has the sampling period of $\SI{2.0}{s}$ and the duration of $\SI{40}{min}$. 
By using the normalized output $s^\mathrm{regd}[k]$ of the test signal at time step $k$, the reference signal $Y_1^\mathrm{ref}[k] $
 for the electric power can be written as 
\begin{align} \label{eq:reference_signal}
 Y_1^\mathrm{ref}[k] = Y_1^\mathrm{nom} +  Y_1^\mathrm{AS}  s^\mathrm{regd}[k]
\end{align}
where $Y_1^\mathrm{nom}$ stands for the nominal power, and $Y_1^\mathrm{AS}$ for the amount of electric power provided for frequency regulation. 
In the following simulations, we set as $Y_1^\mathrm{AS} = \SI{2.0}{MW}$. 
The acceptable range of the pressure variation is set as $[\SI{730}{kPa}, \, \SI{830}{kPa}]$, which is a typical pressure range stipulated in supply agreements of heat suppliers in Japan.

With the above setting, the control input were calculated by solving the optimization problem \eqref{eq:optimization_problem} using the \texttt{quadprog}, a quadratic programming solver in MATLAB. 
The sampling period $T_\mathrm{s}$ was set as $\SI{2.0}{s}$, and the horizon $N$ as $5$. 
The weighting matrices  $\mathsf{Q}_\mathrm{z}$ and  $\mathsf{R}_\mathrm{z}$ were set as  
\begin{align}
 & \mathsf{Q}_z = \mathrm{diag}\{[ [10, \,1, \, 1,\, 1,\, 1]^\top, [10,\,1,\,1,\,1]^\top ]^\top \},  \label{eq:weight} \\
 & \mathsf{R}_z = \mathrm{diag} \{ [0.01, \, 0.01] \}. 
\end{align}
The regularity condition of $\bm{\Phi}$ needed to ensure the validity of the coordinate transformation was checked numerically and satisfied for all the time in the simulations.

\subsection{Computational Issues}

\begin{figure}[!tb]
\begin{minipage}[b]{\linewidth}
  \centering
  \includegraphics[width=.90\linewidth]{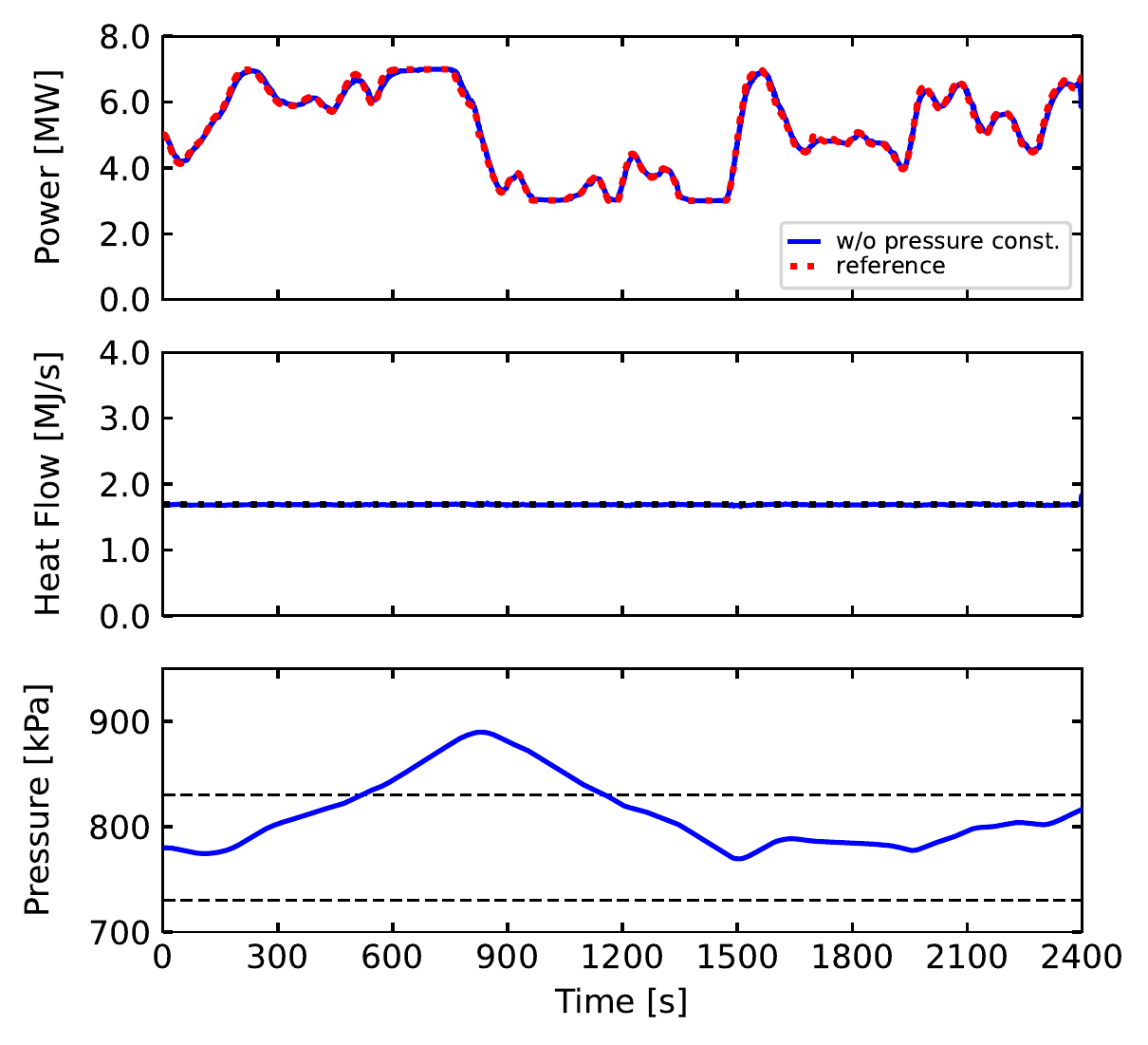}\\[-3mm]
  \subcaption{Outputs} \label{fig:nmpc_outputs}
  \includegraphics[width=.90\linewidth]{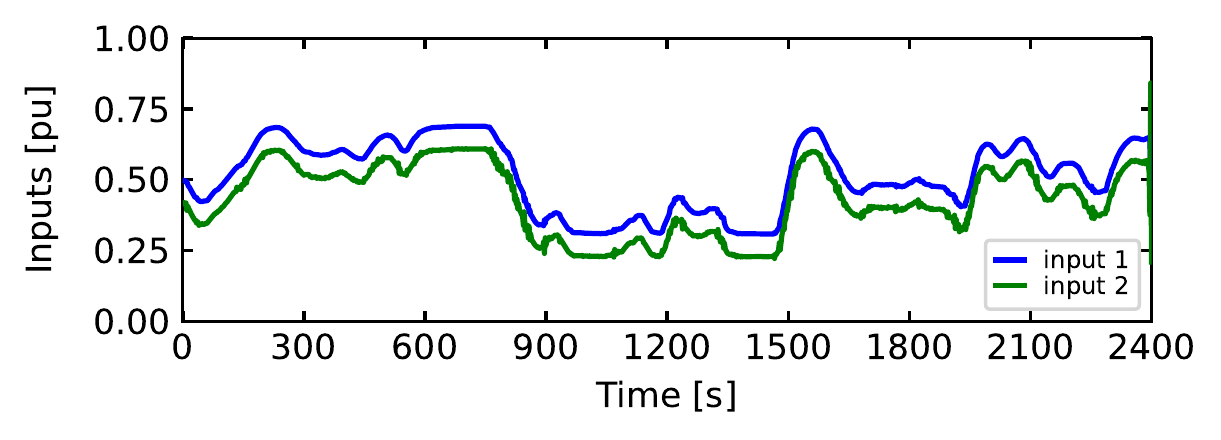}\\[-3mm]
  \subcaption{Inputs}
  \includegraphics[width=.90\linewidth]{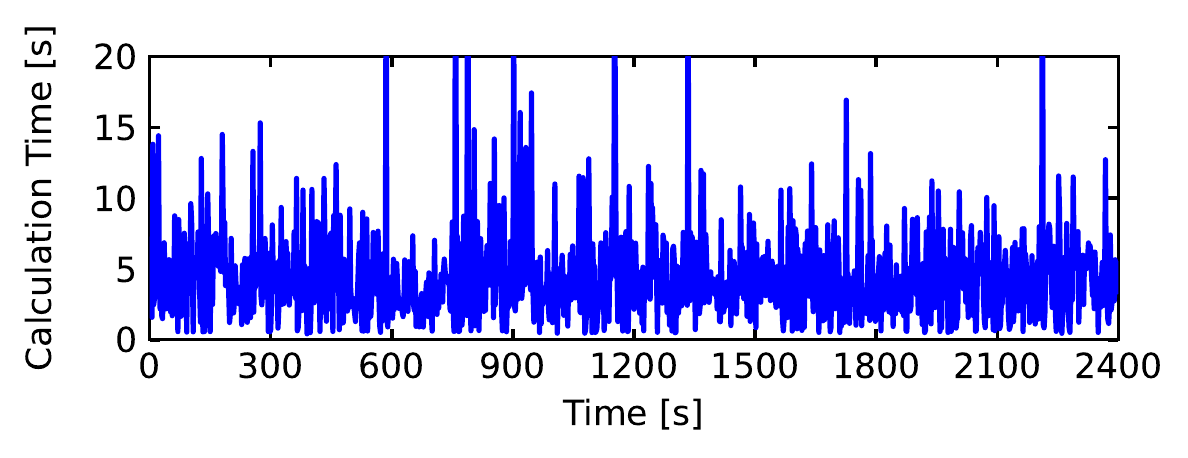}\\[-3mm]
  \subcaption{Computational time} \label{fig_sub1}
\end{minipage}
\caption{Simulation results with a standard nonlinear MPC algorithm. } 
\label{fig:simex1_nmpc}
\end{figure}

\begin{figure}[!tb]
\begin{minipage}[b]{\linewidth}
  \centering
  \includegraphics[width=.90\linewidth]{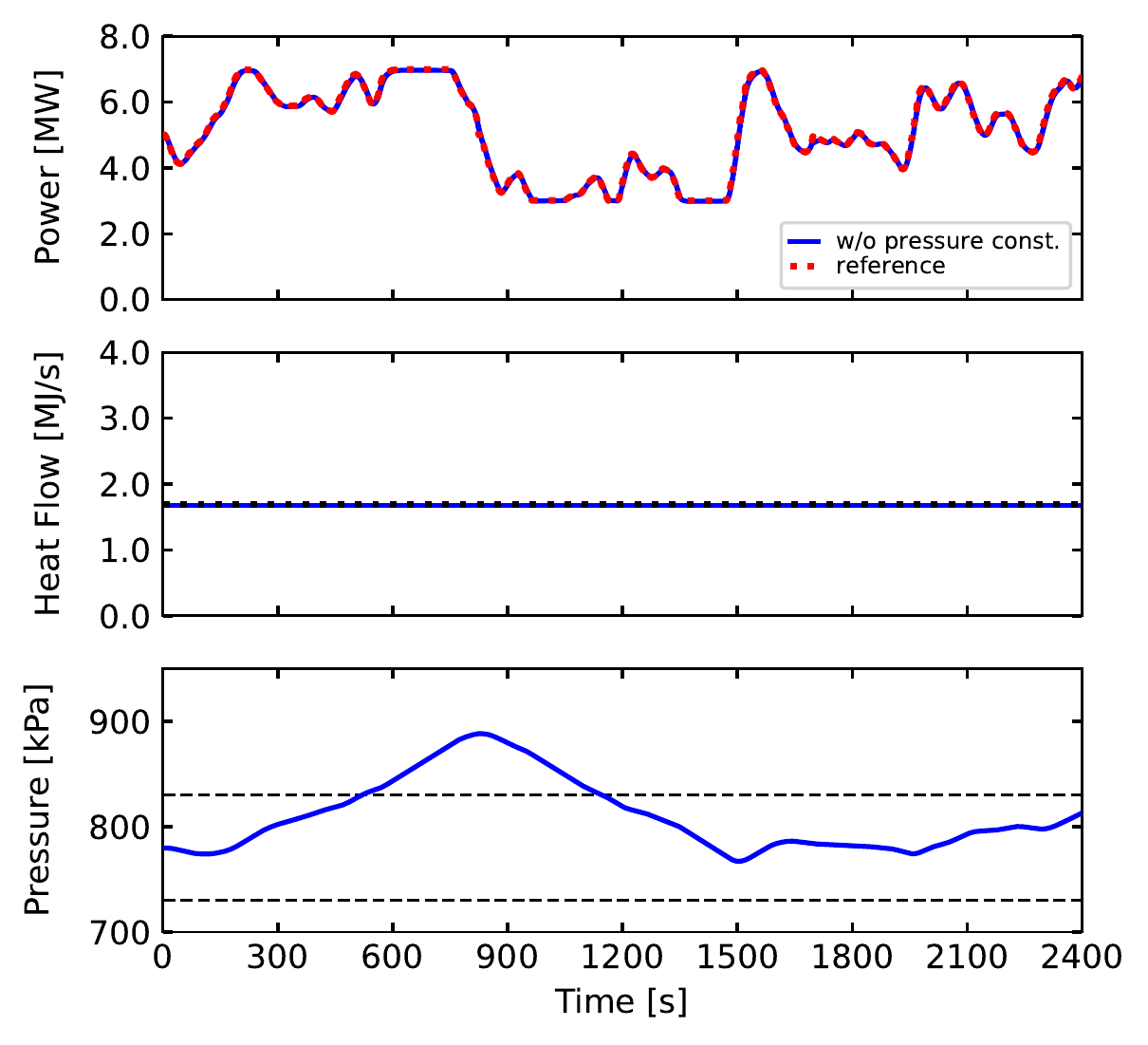}\\[-3mm]
  \subcaption{Outputs} \label{fig_sub1}
  \includegraphics[width=.90\linewidth]{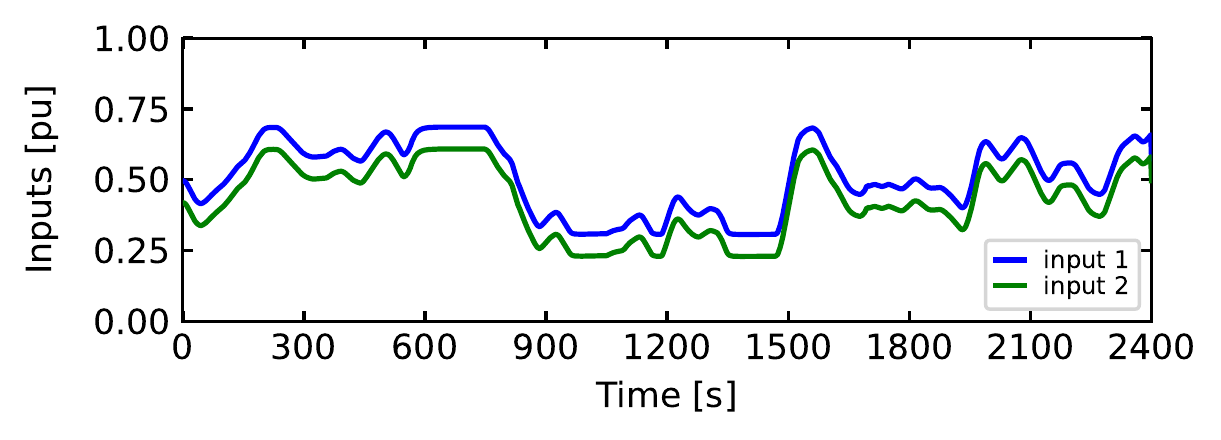}\\[-3mm]
  \subcaption{Inputs}
  \includegraphics[width=.90\linewidth]{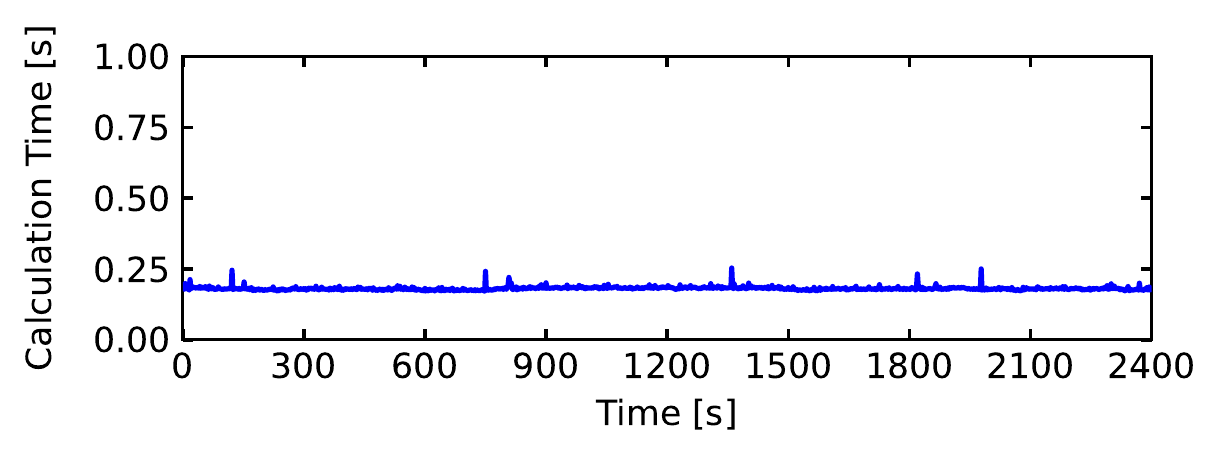}\\[-3mm]
  \subcaption{Computational time} \label{fig_sub1}
\end{minipage}
\caption{Simulation results with the proposed MPC algorithm. } 
\label{fig:simex1_proposed}
\end{figure}

In this subsection, the effectiveness of the proposed MPC 
controller is examined in terms of computational time by a comparison 
with a  standard nonlinear MPC controller \cite{grune11}. 
In the nonlinear MPC controller,  the cost function \eqref{eq:cost_function} is minimized at each time step with 
the interior-point algorithm provided by the \texttt{fmincon} solver, a nonlinear programming solver in MATLAB. 
For this, 
the numerical discretization was performed via integrating the original nonlinear state-space model \eqref{eq:system_equation}, 
which is the default strategy for nonlinear MPC in Model Predictive Control Toolbox in MATLAB \cite{nmpc_matlab}.   
The simulation results with the nonlinear MPC algorithm is shown in \figref{fig:simex1_nmpc}. 
For this simulation, no constraints on the outputs $y_3$ and the inputs $u_1$ and $u_2$ are considered. 
The panel (a) of the figure shows the time courses of the electric output $y_1$, the heat flow $y_2$, and the steam pressure $y_3$, respectively. 
It can be confirmed that the output tracking of $y_1$ and the output regulation of $y_2$ are simultaneously achieved 
by using the two inputs $u_1$ and $u_2$ shown in the panel (b). 
The computational time taken at each time step is shown in the panel (c). 
It takes $\SI{4.4}{s}$ on average and exceeds  $\SI{20}{s}$ frequently while the sampling period $T_\mathrm{s}$ is set as $\SI{2.0}{s}$.

In the proposed MPC framework, the nonlinearities in the original problem are taken care of in two ways:  
1) by using the coordinate transformation $\bm{\Phi}$, we have the cost function \eqref{eq:cost_function} regarding the tracking performance in a quadratic form, which is highly nonlinear in the original coordinate and 2) by using the variational equation for computing the matrices $\mathsf{F}_z[k+l|k], \mathsf{F}_w[k+l|k]$, and $\mathsf{G}_w[k+l|k]$, we have the linear constraints \eqref{eq:constraint_internal_dynamics} and \eqref{eq:constraint_w} regarding the nonlinear internal dynamics, which would be originally described by the map  $\bm{\Psi}$ and some numerical discretization is required. 
As a result, the computation of the original problem with a nonlinear cost function and a numerical discretization is reduced to a quadratic programming problem.
\Figref{fig:simex1_proposed} shows the simulation results with the proposed MPC framework. 
It can be seen that the control performances of the tracking of $y_1$ and the regulation of $y_2$ are similar to the results of the nonlinear MPC. 
Moreover, as shown in the panel (c) of the figure, the computational time taken at each time step is significantly reduced, and it takes $\SI{0.18}{s}$ on average and $\SI{0.25}{s}$ at most and the computation always can be finished within the sampling period  $T_\mathrm{s} = \SI{2.0}{s}$. 
The above results have been obtained in simulation with a $\SI{2.90}{GHz}$ Core i7-based laptop. 

\subsection{Basic Specifications}

\begin{figure}[!tb]
\centering
\includegraphics[width=.90\linewidth]{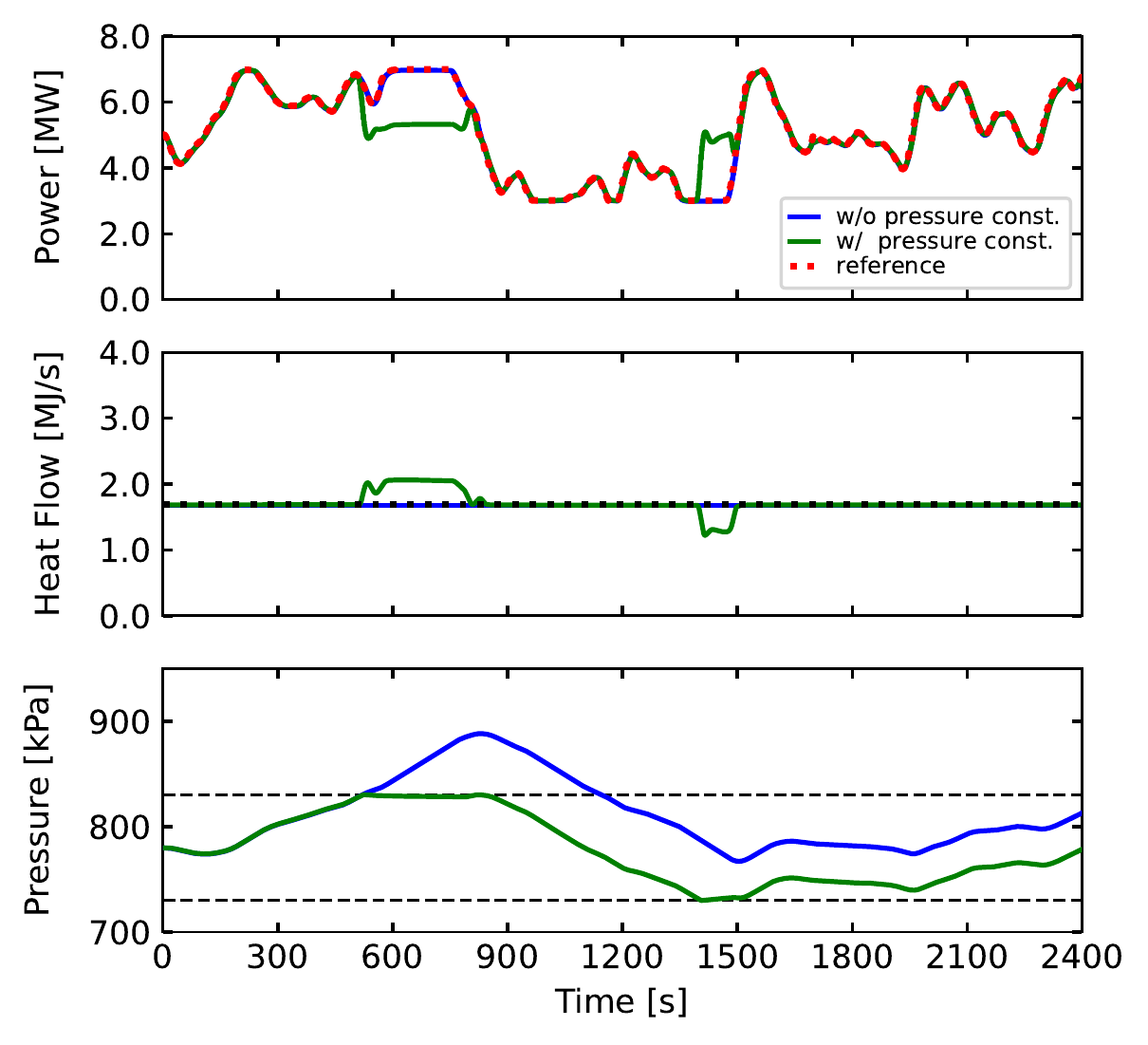} 
\caption{Simulation results with the proposed MPC algorithm with and without the upper and lower limits of the pressure $y_3$. } 
\label{fig:simex2_constraint_output}
\end{figure}

Here we illustrate basic specifications of the proposed MPC controller considering the upper and lower limits of the steam pressure $y_3$. 
\Figref{fig:simex2_constraint_output} shows the time courses of the electric output $y_1$, the heat output $y_2$, and the steam pressure $y_3$. 
%
The \emph{blue} lines in the figure show the responses of the closed-loop system without the constraints on the pressure $y_3$,  
%
and the \emph{green} lines the responses with the constraints. 
It can be observed that the constraint is activated during the period of around $[\SI{500}{s},\, \SI{900}{s}]$ and $[\SI{1400}{s},\, \SI{1500}{s}]$, and the output $y_1$ and $y_2$ deviate from their reference signals during these periods. 
This clearly demonstrates the effectiveness of the designed MPC controller where the energy flows ($y_1$ for  electricity  and $y_2$ for heat) are controlled while maintaining the boundedness of the internal dynamics (pressure $y_3$) for ensuring the supply quality of the steam network.

\begin{figure}[!tb]
\centering
\includegraphics[width=.90\linewidth]{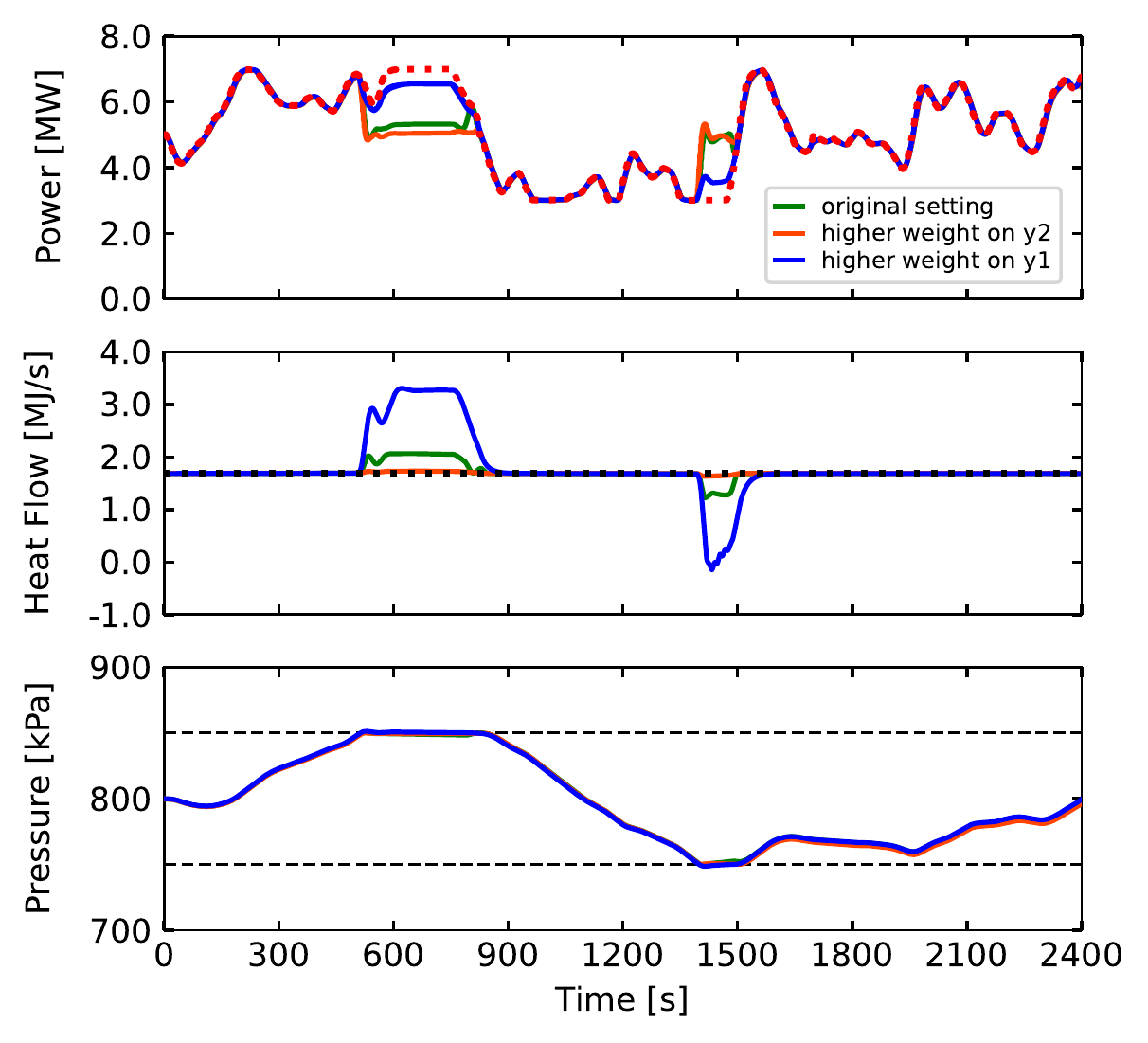} 
\caption{Responses of the closed-loop system for different settings  of the weight $Q_z$. } 
\label{fig:simex2_weight_outputs}
\end{figure}

Furthermore, \figref{fig:simex2_weight_outputs} shows the responses of the closed-loop system where the matrix $\mathsf{Q}_z$ is changed from the setting shown in \eqref{eq:weight}. 
In the figure, the \emph{orange} lines show the simulation results with the setting of 
\begin{align}
 \mathsf{Q}_z = \mathrm{diag}\{[ [10, \,1, \, 1,\, 1,\, 1]^\top, [100,\,1,\,1,\,1]^\top ]^\top \},  \label{eq:weight_Qh}
\end{align}
where the weight on $y_2$ is increased from the original setting, which is shown by the \emph{green} lines in the figure. 
As a result of the increase in the weight on $y_2$, the heat transfer rate $y_2$ becomes closer to the reference value of $Y_2^\mathrm{nom} = \SI{1.69}{MJ/s}$, while degrading the tracking performance of $y_1$. 
%
%
%
%
This simulation result indicates that when the weight on $y_2$ is sufficiently large, the proposed MPC controller exhibits a two-mode behavior as explained below. 
As the first mode of operation, during the period when the constraint \eqref{eq:constraint_w} is not activated, the proposed MPC controller achieves output tracking of $y_1$ and the regulation of $y_2$ to the reference signals, and the steam pressure $y_3$ changes accordingly. 
When the constraint \eqref{eq:constraint_w} is activated, the pressure is kept constant. 
Furthermore, both the outputs $y_1$ and $y_2$ become close to the equilibrium values $y_1 = \SI{5.0}{MW}$ and $y_2 = \SI{1.69}{MJ/s}$, respectively. 
Thus, as the second mode of operation, the proposed MPC controller achieves stabilization of the system to an equilibrium point. 

%
The \emph{blue} lines in \figref{fig:simex2_weight_outputs} shows  the simulation results with the setting of 
\begin{align}
 \mathsf{Q}_z = \mathrm{diag}\{[  [100, \,1, \, 1,\, 1,\, 1]^\top,   [10,\,100,\,100,\,100]^\top ]^\top \},  \label{eq:weight_Qh}
\end{align}
where the weight on $y_1$ is increased from the original setting, and the weights on the derivatives of the output $y_2$ are increased to suppress an abrupt change of the output $y_2$. 
This simulation result indicates that the tracking performance for $y_1$ can be improved if the fluctuation of the heat transfer rate $y_2$ is acceptable, and this trade-off between the tracking of $y_1$ and the regulation of $y_2$ can be easily managed with the proposed method by changing the weight $\mathsf{Q}_\mathrm{z}$. 
%
%

\subsection{Handling of Input Constraints} \label{sec:input_constraints}

\begin{figure}[!tb]
\begin{minipage}[b]{\linewidth}
  \centering
  \includegraphics[width=.89\linewidth]{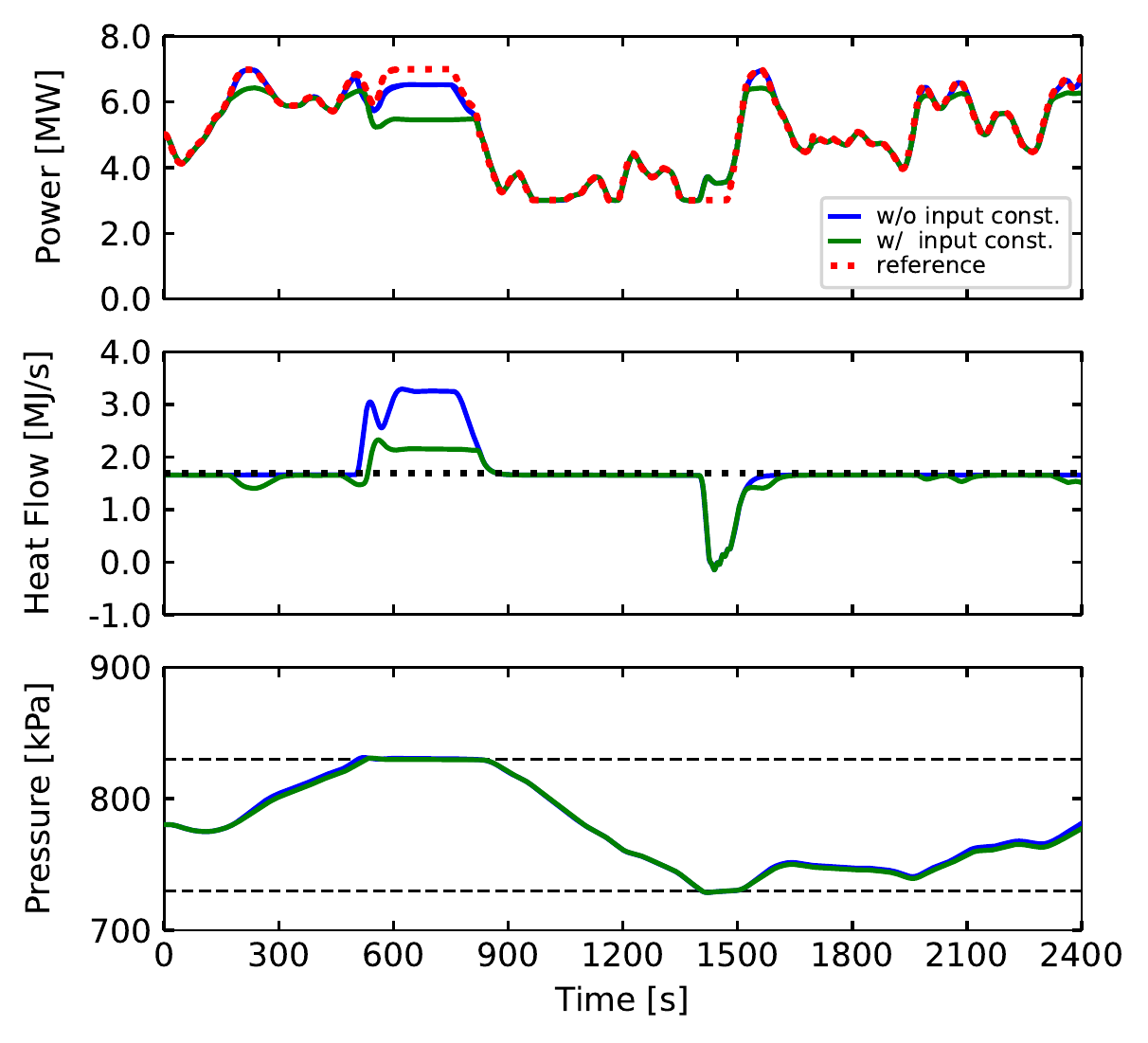}\\[-3mm]
  \subcaption{Outputs} \label{fig:nmpc_outputs}
  \includegraphics[width=.89\linewidth]{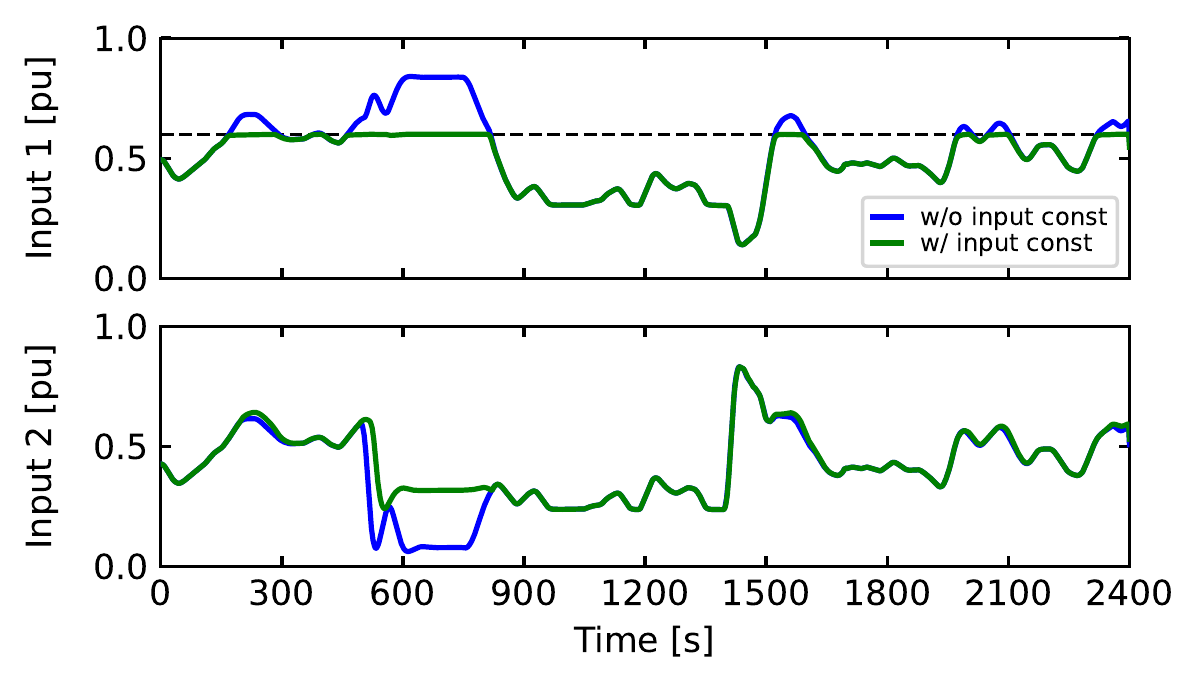}\\[-3mm]
  \subcaption{Inputs}
\end{minipage}
\caption{Responses of the closed-loop system without (shown by \emph{blue}) and with (by \emph{green})  the input constraint $0 \le u_1 \le 0.6$.  } 
\label{fig:simex3_input}
\end{figure}

Finally,  we discuss a case where the input for the site $\# 1$ is limited to $0 \le u_1\le 0.6$, i.e., the maximum limit for $u_1$ is reduced due to a fault or some other reasons. 
To take this into account, the constraints on upper and lower limits of the inputs are converted to a set of constraints on the outputs $y_1$ and $y_2$ represented by \eref{eq:limit_z} by considering  steady states of the system with respect to the states $\bm{\xi}_\mathrm{e}$  and $\bm{\xi}_\mathrm{h}$  given by Eqs.\,\eqref{eq:xi_e_subsystem} and \eqref{eq:xi_h_subsystem}. 
\Figref{fig:simex3_input} shows the responses of the closed-loop system without the input constrains (shown by \emph{blue}) and with the constraints (shown by \emph{green}). 
The setting of the weights in the objective function is the same with that for the simulation shown by the blue lines in \figref{fig:simex2_weight_outputs}. 
As shown in the panel (b) of \figref{fig:simex3_input}, the proposed MPC controller properly works to manage the control objectives 
within  the input constraints.

%
%
%

\begin{figure}[!tb]
 \centering
 \includegraphics[width=0.82\linewidth]{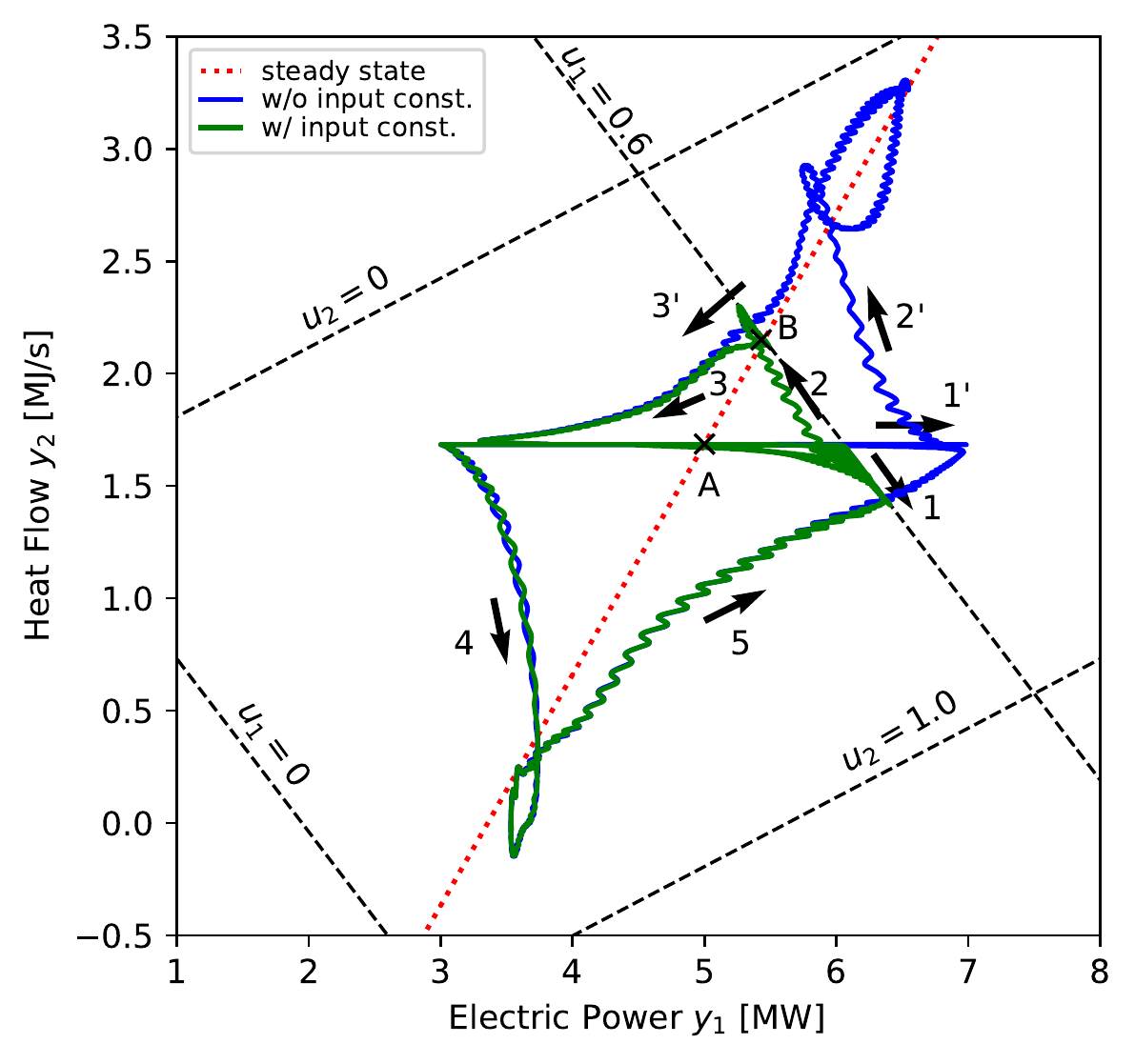} 
 \caption{Two-dimensional portrait of the output pair $(y_1, \, y_2)$.  } 
\label{fig:simex2_refplane}
\end{figure}

%
Furthermore,  
\figref{fig:simex2_refplane} shows a two-dimensional portrait of the output pair $(y_1,\, y_2)$ for each simulation result in \figref{fig:simex3_input}. 
The \emph{black broken} lines in the figure show the constraints \eqref{eq:limit_z} by which upper and lower limits of the inputs $u_1$ and $u_2$ are imposed. 
The point A in the middle of the figure represents the initial condition of the simulation, and the point $(y_1,\, y_2)$ moves horizontally along the line of $y_2= \SI{1.69}{MJ/s}$ if the tracking of $y_1$ and the regulation of $y_2$ are perfectly achieved.  
In this simulation, as shown by the \emph{green} trajectory,  the input constraint $u_1 \le 0.6$ is activated at around $t=\SI{150}{s}$, and then the point $(y_1, y_2)$ starts to move along the line of $u_1 = 0.6$ (shown by the arrow 1), whereas it moves left  if the upper limit on $u_1$ is not imposed (shown by the arrow 1'). 
Then, after the constraint on the pressure $y_3$ is activated at around $t=\SI{500}{s}$, the point $(y_1,\, y_2)$ moves close to the \emph{red} doted line (shown by the arrows $2$ and $2'$). 
This line represents the set of output pairs where an equilibrium point exists and thus the pressure $y_3$ will be kept constant. 
It can be seen that if the input constraint is not imposed, as shown by the \emph{blue} trajectory, the point $(y_1,\, y_2)$ moves around the dotted line, and the output $y_1$ can take the value larger than  $\SI{5.43}{MW}$, which is the value at the point B of the intersection of the lines of $u_1=0.6$ and the steady states. 
However, when the input constraint $u_1 \le 0.6$ is activated, the point  $(y_1,\, y_2)$ approaches to the point B, and the output $y_1$ will be kept at $y_1 = \SI{5.43}{MW}$. 
This is why the tracking performance for $y_1$ was significantly degraded due to the input constraint for the time period of $[\SI{500}{s},\, \SI{850}{s}]$ in \figref{fig:simex3_input}. 
After that, the point $(y_1, \, y_2)$ returns to the line of $y_2 = \SI{1.69}{MJ/s}$ after the constraint on the pressure is deactivated at around $t=\SI{850}{s}$ (shown by the arrows $3$ and $3'$). 
Then, at around $t=\SI{1400}{s}$,  the point $(y_1, \, y_2)$ again moves close to the red doted line when the pressure constraint $y_3 \ge \SI{730}{kPa}$ is activated (shown by the arrow 4). 
Finally, at around $t=\SI{1500}{s}$, with the increase in the reference signal of the output $y_1$, the pressure constraint is deactivated, and the point  $(y_1,\, y_2)$ moves to the upper right direction (shown by the arrow 5). 
As illustrated by this example, the behavior of the proposed MPC controller can be well understood by the portrait of $(y_1\, y_2)$, and the proposed MPC effectively works for managing the complex trade-offs among the three control objectives of the tracking of $y_1$, the regulation of $y_2$, and the boundedness of $y_3$ under the constraints on the inputs. 

\section{Conclusion}

In this paper, we studied a novel control problem to enable participation of CHP units in frequency regulation market while utilizing the thermal inertia of a steam network as storage. 
To ensure safety and supply quality of the steam network, a Model Predictive Control (MPC) framework was presented based on the coordinate transformation to decompose the entire dynamics into the linear outer dynamics related to the tracking performance and the  nonlinear internal dynamics related to the supply quality of the steam network. 
The effectiveness of the proposed MPC framework was demonstrated by numerical simulations of the closed-loop system for a benchmark system based on a practical example of district heating and cooling networks in Japan.

Future directions of this paper include considerations into controller design in the presence of anticipated uncertainties. 
A robust MPC scheme can be implemented without changing the structure of the proposed framework, and its effectiveness will be evaluated in our future contribution.

%

%


\bibliography{draft_hoshino}

\begin{thebibliography}{10}
\providecommand{\url}[1]{#1}
\csname url@samestyle\endcsname
\providecommand{\newblock}{\relax}
\providecommand{\bibinfo}[2]{#2}
\providecommand{\BIBentrySTDinterwordspacing}{\spaceskip=0pt\relax}
\providecommand{\BIBentryALTinterwordstretchfactor}{4}
\providecommand{\BIBentryALTinterwordspacing}{\spaceskip=\fontdimen2\font plus
\BIBentryALTinterwordstretchfactor\fontdimen3\font minus
  \fontdimen4\font\relax}
\providecommand{\BIBforeignlanguage}[2]{{%
\expandafter\ifx\csname l@#1\endcsname\relax
\typeout{** WARNING: IEEEtran.bst: No hyphenation pattern has been}%
\typeout{** loaded for the language `#1'. Using the pattern for}%
\typeout{** the default language instead.}%
\else
\language=\csname l@#1\endcsname
\fi
#2}}
\providecommand{\BIBdecl}{\relax}
\BIBdecl

\bibitem{bollen11}
M.~Bollen and F.~Hassan, \emph{Integration of Distributed Generation in the
  Power System}.\hskip 1em plus 0.5em minus 0.4em\relax John Wiley \& Sons,
  2011.

\bibitem{ostergaard10}
P.~A. {\O}stergaard, ``Regulation strategies of cogeneration of heat and power
  ({CHP}) plants and electricity transit in denmark,'' \emph{Energy}, vol.~35,
  no.~5, pp. 2194--2202, 2010.

\bibitem{korpela17}
T.~Korpela, J.~Kaivosoja, Y.~Majanne, L.~Laakkonen, M.~Nurmoranta, and
  M.~Vilkko, ``Utilization of district heating networks to provide flexibility
  in {CHP} production,'' \emph{Energy Procedia}, vol. 116, pp. 310--319, 2017.

\bibitem{mugnini21}
A.~Mugnini, G.~Comodi, D.~Salvi, and A.~Arteconi, ``Energy flexible {CHP-DHN}
  systems: Unlocking the flexibility in a real plant,'' \emph{Energy Conversion
  and Management: X}, vol.~12, p. 100110, 2021.

\bibitem{du20}
P.~Du, N.~V. Mago, W.~Li, S.~Sharma, Q.~Hu, and T.~Ding, ``New ancillary
  service market for {ERCOT},'' \emph{IEEE Access}, vol.~8, pp.
  178\,391--178\,401, 2020.

\bibitem{abbas21}
A.~O. Abbas and B.~H. Chowdhury, ``Using customer-side resources for
  market-based transmission and distribution level grid services: A review,''
  \emph{International Journal of Electrical Power \& Energy Systems}, vol. 125,
  p. 106480, 2021.

\bibitem{mahani20}
K.~Mahani, M.~A. Jamali, S.~D. Nazemi, and M.~A. Jafari, ``Economic and
  operational evaluation of {PV} and {CHP} combined with energy storage systems
  considering energy and regulation markets,'' in \emph{2020 IEEE Texas Power
  and Energy Conference (TPEC)}, 2020, pp. 1--6.

\bibitem{gao22}
S.~Gao, J.~Jurasz, H.~Li, E.~Corsetti, and J.~Yan, ``Potential benefits from
  participating in day-ahead and regulation markets for {CHPs},'' \emph{Applied
  Energy}, vol. 306, p. 117974, 2022.

\bibitem{hennessy19}
J.~Hennessy, H.~Li, F.~Wallin, and E.~Thorin, ``Flexibility in thermal grids: a
  review of short-term storage in district heating distribution networks,''
  \emph{Energy Procedia}, vol. 158, pp. 2430--2434, 2019.

\bibitem{parisio17}
A.~Parisio, C.~Wiezorek, T.~Kynt\:{a}j\:{a}, J.~Elo, K.~Strunz, and K.~H.
  Johansson, ``Cooperative {MPC}-based energy management for networked
  microgrids,'' \emph{IEEE Transactions on Smart Grid}, vol.~8, no.~6, pp.
  3066--3074, 2017.

\bibitem{cao19}
Y.~Cao, W.~Wei, L.~Wu, S.~Mei, M.~Shahidehpour, and Z.~Li, ``Decentralized
  operation of interdependent power distribution network and district heating
  network: A market-driven approach,'' \emph{IEEE Transactions on Smart Grid},
  vol.~10, no.~5, pp. 5374--5385, 2019.

\bibitem{zhang19}
C.~Zhang, Y.~Xu, Z.~Li, and Z.~Y. Dong, ``Robustly coordinated operation of a
  multi-energy microgrid with flexible electric and thermal loads,'' \emph{IEEE
  Transactions on Smart Grid}, vol.~10, no.~3, pp. 2765--2775, 2019.

\bibitem{li22}
Z.~Li, L.~Wu, Y.~Xu, S.~Moazeni, and Z.~Tang, ``Multi-stage real-time operation
  of a multi-energy microgrid with electrical and thermal energy storage
  assets: A data-driven mpc-adp approach,'' \emph{IEEE Transactions on Smart
  Grid}, vol.~13, no.~1, pp. 213--226, 2022.

\bibitem{good19}
N.~Good and P.~Mancarella, ``Flexibility in multi-energy communities with
  electrical and thermal storage: A stochastic, robust approach for
  multi-service demand response,'' \emph{IEEE Transactions on Smart Grid},
  vol.~10, no.~1, pp. 503--513, 2019.

\bibitem{martinez19}
E.~A. Mart\'{i}nez Cese\~{n}a and P.~Mancarella, ``Energy systems integration
  in smart districts: Robust optimisation of multi-energy flows in integrated
  electricity, heat and gas networks,'' \emph{IEEE Transactions on Smart Grid},
  vol.~10, no.~1, pp. 1122--1131, 2019.

\bibitem{long19}
S.~Long, O.~Marjanovic, and A.~Parisio, ``Generalised control-oriented
  modelling framework for multi-energy systems,'' \emph{Applied Energy}, vol.
  235, pp. 320--331, 2019.

\bibitem{lund14}
H.~Lund, S.~Werner, R.~Wiltshire, S.~Svendsen, J.~E. Thorsen, F.~Hvelplund, and
  B.~V. Mathiesen, ``4th generation district heating ({4GDH}): Integrating
  smart thermal grids into future sustainable energy systems,'' \emph{Energy},
  vol.~68, pp. 1--11, 2014.

\bibitem{werner17}
S.~Werner, ``International review of district heating and cooling,''
  \emph{Energy}, vol. 137, pp. 617--631, 2017.

\bibitem{suzuki21}
N.~Suzuki, T.~Shirane, T.~Morimura, and H.~Nitta, ``Construction of underground
  service tunnels as part of disaster management,'' \emph{{IOP} Conference
  Series: Earth and Environmental Science}, vol. 703, no.~1, p. 012040, 2021.

\bibitem{zeng21}
X.~Zeng, X.~Lin, W.~Zhong, L.~Wang, F.~Kong, L.~Li, and G.~Pan, ``Model-based
  electric regulating valve design in looped steam heating system user-side
  temperature control,'' in \emph{2021 IEEE Sustainable Power and Energy
  Conference (iSPEC)}, 2021, pp. 86--94.

\bibitem{omalley13}
M.~O'Malley and B.~Kroposki, ``Energy comes together: The integration of all
  systems,'' \emph{IEEE Power \& Energy Magazine}, vol.~11, no.~5, pp. 18--23,
  Sept 2013.

\bibitem{wu16}
J.~Wu, J.~Yan, H.~Jia, N.~Hatziargyriou, N.~Djilali, and H.~Sun, ``Integrated
  energy systems,'' \emph{Applied Energy}, vol. 167, pp. 155--157, 2016.

\bibitem{mancarella14}
P.~Mancarella, ``{MES} (multi-energy systems): An overview of concepts and
  evaluation models,'' \emph{Energy}, vol.~65, pp. 1--17, 2014.

\bibitem{chertkov20}
M.~Chertkov and G.~Andersson, ``Multienergy systems,'' \emph{Proceedings of the
  IEEE}, vol. 108, no.~9, pp. 1387--1391, 2020.

\bibitem{geidl07:IEEETPS}
M.~Geidl and G.~Andersson, ``Optimal power flow of multiple energy carriers,''
  \emph{IEEE Transactions on Power Systems}, vol.~22, no.~1, pp. 145--155,
  2007.

\bibitem{geidl07:PESMAG}
M.~Geidl, G.~Koeppel, P.~Favre-Perrod, B.~Klockl, G.~Andersson, and
  K.~Frohlich, ``Energy hubs for the future,'' \emph{IEEE Power and Energy
  Magazine}, vol.~5, no.~1, pp. 24--30, 2007.

\bibitem{arnold09}
M.~Arnold, R.~R. Negenborn, G.~Andersson, and B.~De~Schutter, ``Model-based
  predictive control applied to multi-carrier energy systems,'' in \emph{2009
  IEEE Power \& Energy Society General Meeting}, 2009, pp. 1--8.

\bibitem{pan16}
Z.~Pan, Q.~Guo, and H.~Sun, ``Interactions of district electricity and heating
  systems considering time-scale characteristics based on quasi-steady
  multi-energy flow,'' \emph{Applied Energy}, vol. 167, pp. 230--243, 2016.

\bibitem{pan19}
Z.~Pan, J.~Wu, H.~Sun, Q.~Guo, and M.~Abeysekera, ``Quasi-dynamic interactions
  and security control of integrated electricity and heating systems in normal
  operations,'' \emph{CSEE Journal of Power and Energy Systems}, vol.~5, no.~1,
  pp. 120--129, 2019.

\bibitem{hoshino14:nolta}
H.~Hoshino, Y.~Susuki, and T.~Hikihara, ``A nonlinear dynamical model of
  two-sites electricity and heat supply system,'' in \emph{2014 International
  Symposium on Nonlinear Theory and its Applications}, 2014, pp. 482--485.

\bibitem{hoshino16}
------, ``{A Lumped-Parameter Model of Multiscale Dynamics in Steam Supply
  Systems},'' \emph{Journal of Computational and Nonlinear Dynamics}, vol.~11,
  no.~6, 2016, 061018.

\bibitem{hoshino19}
H.~Hoshino, Y.~Susuki, T.~J. Koo, and T.~Hikihara, ``{Structural Analysis and
  Control of a Model of Two-Site Electricity and Heat Supply},'' \emph{Journal
  of Dynamic Systems, Measurement, and Control}, vol. 141, no.~10, 2019,
  101004.

\bibitem{kundur94}
P.~Kundur, \emph{Power System Stability and Control}.\hskip 1em plus 0.5em
  minus 0.4em\relax McGraw-Hill, 1994.

\bibitem{pjm_manual12}
PJM, ``{PJM} manual 12: Balancing operations,'' 2022,
  \url{https://www.pjm.com/~/media/documents/manuals/m12.ashx} [Accessed:
  August 19, 2022].

\bibitem{isidori95}
A.~Isidori, \emph{Nonlinear Control Systems}, 3rd~ed.\hskip 1em plus 0.5em
  minus 0.4em\relax Springer-Verlag, 1995.

\bibitem{sastry99}
S.~Sastry, \emph{Nonlinear Systems: Analysis, Stability, and Control}.\hskip
  1em plus 0.5em minus 0.4em\relax Springer-Verlag, 1999.

\bibitem{li04}
W.~Li and E.~Todorov, ``Iterative linear quadratic regulator design for
  nonlinear biological movement systems.'' in \emph{1st Int. Conf. Informatics
  in Control, Automation and Robotics}, 2004, pp. 222--229.

\bibitem{todorov05}
E.~Todorov and W.~Li, ``A generalized iterative {LQG} method for
  locally-optimal feedback control of constrained nonlinear stochastic
  systems,'' in \emph{Proceedings of the 2005 American Control Conference},
  2005, pp. 300--306.

\bibitem{hatziargyriou07}
N.~Hatziargyriou, H.~Asano, R.~Iravani, and C.~Marnay, ``Microgrids,''
  \emph{IEEE Power and Energy Magazine}, vol.~5, no.~4, pp. 78--94, 2007.

\bibitem{araposthatis81}
A.~Araposthatis, S.~Sastry, and P.~Varaiya, ``Analysis of power-flow
  equation,'' \emph{International Journal of Electrical Power \& Energy
  Systems}, vol.~3, no.~3, pp. 115--126, 1981.

\bibitem{arapostathis83}
A.~Arapostathis and P.~Varaiya, ``Behaviour of three-node power networks,''
  \emph{International Journal of Electrical Power \& Energy Systems}, vol.~5,
  no.~1, pp. 22--30, 1983.

\bibitem{hasegawa99}
Y.~Hasegawa and Y.~Ueda, ``Global basin structure of attraction of two degrees
  of freedom swing equation system,'' \emph{International Journal of
  Bifurcation and Chaos}, vol.~09, no.~08, pp. 1549--1569, 1999.

\bibitem{nedo_kobe18}
``News release: Nedo conducts world's first technology demonstration for
  hydrogen-fueled cogeneration system in urban areas,'' Available:
  \url{http://www.nedo.go.jp/english/news/AA5en_100348.html} [Accessed:
  15-Mar-2023].

\bibitem{zhong15}
W.~Zhong, H.~Feng, X.~Wang, D.~Wu, M.~Xue, and J.~Wang, ``Online hydraulic
  calculation and operation optimization of industrial steam heating networks
  considering heat dissipation in pipes,'' \emph{Energy}, vol.~87, pp.
  566--577, 2015.

\bibitem{nevistic94}
V.~Nevistic and L.~Del~Re, ``Feasible suboptimal model predictive control for
  linear plants with state dependent constraints,'' in \emph{Proceedings of
  1994 American Control Conference}, vol.~3, 1994, pp. 2862--2866.

\bibitem{nevistic96}
V.~Nevisti\'{c} and M.~Morari, ``Robustness of mpc-based schemes for
  constrained control of nonlinear systems,'' in \emph{Proceedings of the 13th
  World Congress of IFAC}, vol.~29, no.~1, 1996, pp. 5823--5828.

\bibitem{margellos10}
K.~Margellos and J.~Lygeros, ``A simulation based {MPC} technique for feedback
  linearizable systems with input constraints,'' in \emph{49th IEEE Conference
  on Decision and Control}, 2010, pp. 7539--7544.

\bibitem{simon13}
D.~Simon, J.~L\:{o}fberg, and T.~Glad, ``Nonlinear model predictive control
  using feedback linearization and local inner convex constraint
  approximations,'' in \emph{2013 European Control Conference}, 2013, pp.
  2056--2061.

\bibitem{schnelle15}
F.~Schnelle and P.~Eberhard, ``Constraint mapping in a feedback
  linearization/{MPC} scheme for trajectory tracking of underactuated multibody
  systems,'' in \emph{5th IFAC Conference on Nonlinear Model Predictive
  Control}, 2015, pp. 446--451.

\bibitem{gionfra16}
N.~Gionfra, H.~Siguerdidjane, G.~Sandou, D.~Faille, and P.~Loevenbruck,
  ``Combined feedback linearization and mpc for wind turbine power tracking,''
  in \emph{2016 IEEE Conference on Control Applications (CCA)}, 2016, pp.
  52--57.

\bibitem{quan21}
S.~Quan, Y.-X. Wang, X.~Xiao, H.~He, and F.~Sun, ``Feedback linearization-based
  mimo model predictive control with defined pseudo-reference for hydrogen
  regulation of automotive fuel cells,'' \emph{Applied Energy}, vol. 293, p.
  116919, 2021.

\bibitem{kong23}
X.~Kong, M.~A. Abdelbaky, X.~Liu, and K.~Y. Lee, ``Stable feedback
  linearization-based economic {MPC} scheme for thermal power plant,''
  \emph{Energy}, vol. 268, p. 126658, 2023.

\bibitem{parker89}
T.~Parker and L.~Chua, \emph{Practical Numerical Algorithms for Chaotic
  Systems}.\hskip 1em plus 0.5em minus 0.4em\relax Springer, 1989.

\bibitem{grune11}
L.~Gr\"{u}ne and J.~Pannek, \emph{Nonlinear Model Predictive Control: Theory
  and Algorithms}.\hskip 1em plus 0.5em minus 0.4em\relax Springer-Verlag,
  2011.

\bibitem{nmpc_matlab}
MathWorks, ``Nonlinear {MPC} design,'' Available:
  \url{https://www.mathworks.com/help/mpc/nonlinear-mpc-design.html} [Accedded:
  May 7, 2023].

\end{thebibliography}
\bibliographystyle{IEEEtran}

\end{document}